%% file: main.tex
\documentclass[sigconf]{acmart}

\usepackage{booktabs,makecell}
\usepackage{graphicx}
\usepackage{geometry}

\usepackage{subcaption}
\usepackage{array}
\usepackage{multirow}
\usepackage{todonotes}
\usepackage{enumitem}
\usepackage{amsmath}
\usepackage{subcaption}
\usepackage[utf8]{inputenc} 
\usepackage[T1]{fontenc}    
\usepackage{hyperref}       
\usepackage{url}            
\usepackage{booktabs}       
\usepackage{amsfonts}       
\usepackage{nicefrac}       
\usepackage{microtype}      
\usepackage{xcolor}         
\usepackage{dsfont}

\usepackage{bbding}
\usepackage{pifont}
\usepackage{wasysym}

\usepackage{makecell}
\usepackage{newfloat}
\usepackage{colortbl}
\definecolor{myyellow}{rgb}{1,1, 0.6}
\definecolor{myorange}{rgb}{1, 0.8, 0.6}
\definecolor{myred}{rgb}{1, 0.6, 0.6}
\definecolor{second}{HTML}{FFDAB9}
\definecolor{best}{HTML}{FFC1C1}
\usepackage{stackrel}
\usepackage{enumitem}
\usepackage{mathtools}
\usepackage{makecell}

\usepackage{wrapfig} 
\usepackage{wrapfig,lipsum,booktabs}
\usepackage{ulem}

\copyrightyear{2024}
\acmYear{2024}
\setcopyright{acmlicensed}\acmConference[WWW '24]{Proceedings of the ACM Web Conference 2024}{May 13--17, 2024}{Singapore, Singapore}
\acmBooktitle{Proceedings of the ACM Web Conference 2024 (WWW '24), May 13--17, 2024, Singapore, Singapore}
\acmDOI{10.1145/3589334.3645316}
\acmISBN{979-8-4007-0171-9/24/05}

\renewcommand{\todo}[1]{\iffalse #1 \fi{\color{blue} \textbf{[TODO]}}}

\newcommand{\leftrarrows}{\mathrel{\raise.9ex\hbox{\oalign{%
  $\scriptstyle\leftarrow$\cr
  \vrule width0pt height.5ex$\hfil\scriptstyle\relbar$\cr}}}}
\newcommand{\lrightarrows}{\mathrel{\raise.9ex\hbox{\oalign{%
  $\scriptstyle\relbar$\hfil\cr
  $\scriptstyle\vrule width0pt height.5ex\smash\rightarrow$\cr}}}}
\newcommand{\Rrelbar}{\mathrel{\raise.9ex\hbox{\oalign{%
  $\scriptstyle\relbar$\cr
  \vrule width0pt height.5ex$\scriptstyle\relbar$}}}}
\newcommand{\longleftrightarrows}{\leftrarrows\joinrel\Rrelbar\joinrel\lrightarrows}

\usepackage{eqparbox}

\newcommand*{\method}{IntellectReq}

\title{Intelligent Model Update Strategy for Sequential Recommendation}


\author{Zheqi Lv}
\affiliation{%
  \institution{Zhejiang University}
  \city{Hangzhou}
  \country{China}}
\email{zheqilv@zju.edu.cn}

\author{Wenqiao Zhang}
\affiliation{%
  \institution{Zhejiang University}
  \city{Hangzhou}
  \country{China}}
\email{wenqiaozhang@zju.edu.cn}

\author{Zhengyu Chen}
\affiliation{%
 \institution{Zhejiang University}
 \city{Hangzhou}
 \country{China}}
\email{chenzhengyu@zju.edu.cn}

\author{Shengyu Zhang}
\affiliation{
  \institution{Zhejiang University}
  \city{Hangzhou}
  \country{China}
  }
\authornote{Corresponding authors.}
\email{sy_zhang@zju.edu.cn}

\author{Kun Kuang}
\affiliation{%
  \institution{Zhejiang University}
  \city{Hangzhou}
  \country{China}}
\authornotemark[1]
\email{kunkuang@zju.edu.cn}

\renewcommand{\shortauthors}{Zheqi Lv et al.}

\begin{document}

\input{tex/0abstract}
\maketitle
\input{tex/1introduction}
\input{tex/2related_work}
\input{tex/3method}
\input{tex/4experiments}

\input{tex/5conclusion}
\input{tex/7acknowledge}

\clearpage

{\small
\bibliographystyle{ACM-Reference-Format}
\bibliography{reference}
}

\input{tex/99appendix}
\end{document}

%% file: tex/0abstract.tex
\begin{abstract}
\label{sec:abstract}

Modern online platforms are increasingly employing recommendation systems to address information overload and improve user engagement.
There is an evolving paradigm in this research field that recommendation network learning occurs both on the cloud and on edges with knowledge transfer in between (\textit{i.e.}, edge-cloud collaboration). Recent works push this field further by enabling edge-specific context-aware adaptivity, where model parameters are updated in real-time based on incoming on-edge data. However, we argue that frequent data exchanges between the cloud and edges often lead to inefficiency and waste of communication/computation resources, as considerable parameter updates might be redundant. To investigate this problem, we introduce \textbf{\underline{Intell}}igent \textbf{\underline{E}}dge-\textbf{\underline{C}}loud Parame\textbf{\underline{t}}er \textbf{\underline{Req}}uest Model~(\textbf{IntellectReq}).
IntellectReq is designed to operate on edge, evaluating the cost-benefit landscape of parameter requests with minimal computation and communication overhead. We formulate this as a novel learning task, aimed at the detection of out-of-distribution data, thereby fine-tuning adaptive communication strategies. Further, we employ statistical mapping techniques to convert real-time user behavior into a normal distribution, thereby employing multi-sample outputs to quantify the model’s uncertainty and thus its generalization capabilities. Rigorous empirical validation on three widely-adopted benchmarks evaluates our approach, evidencing a marked improvement in the efficiency and generalizability of edge-cloud collaborative and dynamic recommendation systems.

\end{abstract}

\keywords{Edge-Cloud Collaboration, Distribution Shift, Mis-Recommendation Detection, Out-of-Domain Detection, Sequential Recommendation}


%% file: tex/1introduction.tex
\begin{figure*}[t]
  \centering
\includegraphics[width=0.88\linewidth]{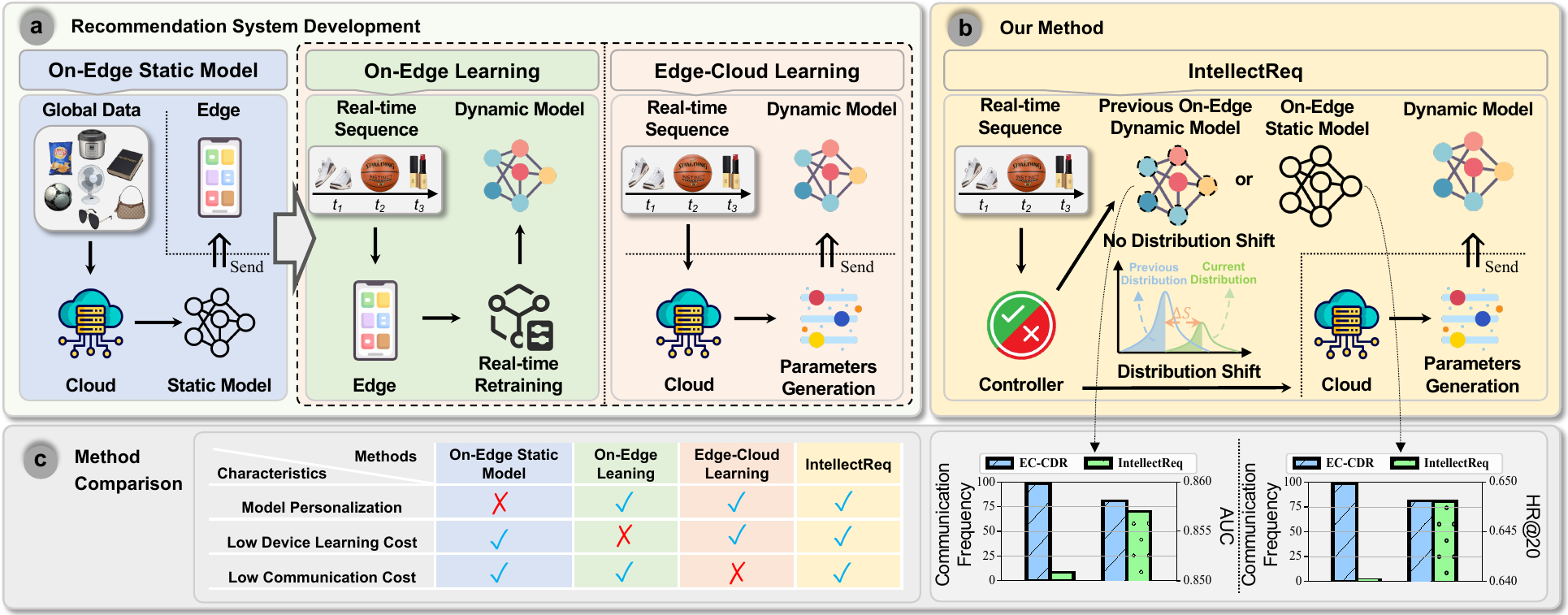}
\vspace{-0.35cm}
  \caption{(a) outlines the evolution of recommendation systems from On-Edge Static Model to Edge-Cloud Collaborative and Dynamic Recommendation (EC-CDR). (b) overviews our IntellectReq. (c) compares the three methods and our IntellectReq (\uline{\textbf{\texttt{Communication Frequency 10\% (IntellectReq)~$\ll$~100\% (EC-CDR), AUC: 0.8562 (IntellectReq) $\approx$ 0.8581 (EC-CDR)}} or \textbf{\texttt{Communication Frequency 3\% (IntellectReq)~$\ll$~100\% (EC-CDR), HR@20: 0.6478 (IntellectReq) $\approx$ 0.6478 (EC-CDR)}}})}
  \label{fig:motivation1}
\vspace{-0.5cm}
\end{figure*}
\section{Introduction}

With the rapid development of e-commerce and social media platforms, recommendation systems~\cite{ref:gru4rec,ref:sasrec,zhangsy2023personalized,lv2022personalizing,zhang2024mitigating} have become indispensable tools in people's daily life. They can be recognized as various forms depending on industries, like product suggestions on online e-commerce websites, (\textit{e.g.}, Amazon and Taobao) or playlist generators for video and music services (\textit{e.g.}, YouTube, Netflix, and Spotify). Among them, one of the classical recommendation systems in the industry prefers to trains a universal model with static parameters  on a powerful cloud conditioned on rich data collected from different edges, and then perform edge inference for all users, such as \textit{e.g.}, DIN~\cite{ref:din}, SASRec~\cite{ref:sasrec}, and GRU4Rec~\cite{ref:gru4rec}. In the first model presented in Figure~\ref{fig:motivation1}, this cloud-based static model allows users to share a centralized model, enabling real-time inference across all edges. However, it does not take advantage of the personalized recommendation patterns specific to each edge due to the shift in data distribution between the cloud and edge. As we all know, the shift in the distribution of test data compared to training data will reduce the performance of the model~\cite{chen2021multi,chen2022ba,zhang2021magic,zhang2022boostmis,zhang2023learning,zhu2023universal,DBLP:conf/mm/ZhuL0HWK0W23,DBLP:conf/kdd/TongYZZZWK23,zhang2022fairness,zhang2023federated,zhang2024revisiting}.
 
To address this issue, existing solutions can be broadly classified into two categories: (i) \textit{On-Edge Learning}: It improve personalization by on-edge learning with the second method depicted in Figure~\ref{fig:motivation1}(a), based on the on-edge static model. Techniques such as distillation~\cite{ref:disitll} and fine-tuning~\cite{ref:finetuning} can mitigate the discrepancy between edge and cloud distributions through re-training at the edge. However, retraining at the edge involves a significant amount of computation, particularly in backpropagation. The sudden drop in real-time performance also reduces its practicality. (ii) \textit{Edge-Cloud Collaboration}~\cite{ref:edge_cloud,ref:edge_cloud2}: It leverages the edge-cloud collaboration to efficiently update the parameters of the edge-model according to on-edge real-time data distribution~\cite{ref:duet,ref:apg_rs1}. Recent advancements have introduced a technique known as adaptive parameter generation~\cite{ref:apg_rs1,ref:duet} (shown as the third method in Figure~\ref{fig:motivation1}(a)), which facilitating model personalization without additional on-edge computational cost. This method specifically utilizes a pre-trained hypernetwork~\cite{ref:hypernetwork_pioneering1} to convert the user's real-time click sequence into adaptive parameters through forward propagation. These parameters then be updated to the edge model, allowing it to better fit real-time data distribution for swift personalization of recommendations. This method, termed the \textbf{\underline{E}}dge-\textbf{\underline{C}}loud \textbf{\underline{C}}ollaborative  and \textbf{\underline{D}}ynamic \textbf{\underline{R}}ecommendation~(\textbf{EC-CDR}), offers tailored recommendation models across various on-edge distribution. 

\begin{sloppypar}
EC-CDR faces deployment challenges in real-world settings due to two key issues:
(i) \textbf{High Request Frequency.} Updating EC-CDR model parameters through edge-cloud communication after a user clicks a new item causes a surge in concurrent cloud requests from multiple edges in industrial settings. This problem worsens in unstable networks, limiting EC-CDR's efficiency due to communication and network constraints. (ii) \textbf{Low Communication Revenue.} When the latest real-time data is the same as, or closely related to, the distribution used previously to update model parameters, communication from edge to cloud is unnecessary. That is, the moment of distribution shift does not always coincide with the timing of model updates at the edge. Unnecessary communication between cloud and edge can lead to low efficiency in communication resource utilization.

\end{sloppypar}
\begin{wrapfigure}[10]{l}{0.278\textwidth}
\centering
\vspace{-0.5cm}
    \begin{center}
    \includegraphics[width=0.24\textwidth]{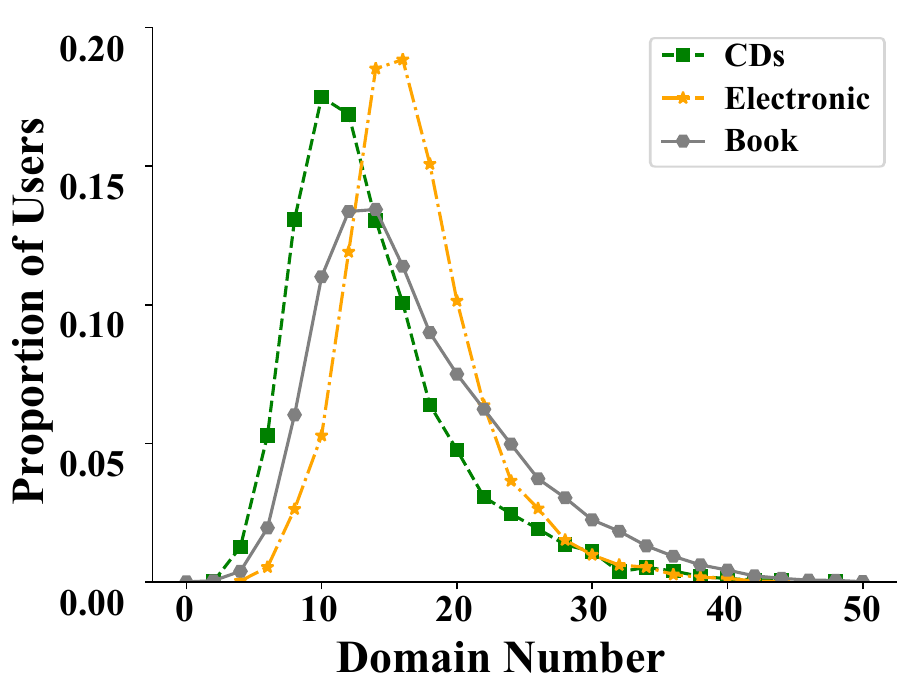}
    \vspace{-0.4cm}
    \caption{Domain numbers of users.}
    \label{fig:distribution_shift}
    \end{center}
\end{wrapfigure}
\begin{sloppypar}
To address EC-CDR's communication issues, we analyzed users' click classes (viewed as domains) on the edge. As shown in Figure~\ref{fig:distribution_shift}, by collecting item embedding vectors from user clicks across three datasets and classifying them into 50 domains, we found that users typically engage with only 10 to 15 domains. This repeated behavior indicates a failure of EC-CDR to recognize shifts in data distribution on the edge, resulting in frequent dynamic parameter requests and high communication overhead.
\end{sloppypar}

\begin{sloppypar}
Based on the insights discussed earlier, our primary optimization goal is to minimize unnecessary communications, aiming for a highly efficient EC-CDR system. To achieve this, we design IntellectReq for deployment on the edge, tasked with assessing the necessity of requests with minimal resource usage. This strategy significantly boosts efficient communication in EC-CDR. IntellectReq is operationalized through the development of the Mis-Recommendation Detector (MRD) and Distribution Mapper (DM).
The MRD is engineered to predict the likelihood of edge recommendation models making incorrect recommendations, termed as Mis-Recommendations. It accomplishes this by learning to map current data and previous data which is used to update the last model to mis-recommendation labels. Moreover, MRD translates these predictions into the potential revenue from updating the edge model, thus maximizing revenue within any communication budget and ensuring the model's optimal performance. The DM is designed to allow the model to detect potential shifts in data distribution and assess the model's uncertainty in interpreting real-time data, which in turn, augments the capabilities of the MRD module. It comprises three components: a prior network, a posterior network, and a next-item prediction network, with the last serving as DM's backbone. During the training phase, data features are extracted through both prior and posterior networks, using label-provided prior information to enhance training efficiency. In the inference stage, the posterior network is utilized for feature extraction. By evaluating the model's uncertainty in processing real-time data—achieved by mapping this data to a normal distribution—DM significantly improves MRD's prediction accuracy. The conventional recommendation datasets prove inadequate for these tasks. Therefore, we have restructured these datasets into a new MRD dataset, eliminating the need for extra annotations. This restructuring process provides essential supervisory data for training our MRD and DM models, ensuring their effectiveness in the EC-CDR system.
\end{sloppypar}

To summarize, our contributions are four-fold:

\begin{sloppypar}
\begin{itemize}[itemsep=1pt,topsep=2pt,leftmargin=20pt]
\item We are the first to point out and introduce IntellectReq to address the issues of high communication frequency and low communication revenue in EC-CDR, a method that improves edge recommendation models to SOTA performance and achieve personalized updates without retraining. 
\item We designed IntellectReq and developed both a Mis-Recommendation Detector (MRD) and a Distribution Mapper (DM) to instantiate IntellectReq. IntellectReq can quantify changes in the data distribution on the edge, and based on the actual communication budget or cloud computing budget, it can determine which edge models need to be updated.
\item  We construct Mis-Recommendation datasets from existing recommendation datasets, as current datasets are not suitable for training IntellectReq, thereby enabling its training without requiring additional manual annotations.
\item We evaluate our method with extensive experiments. Experiments demonstrate that IntellectReq can achieve high revenue under any edge-cloud communication budget.
\end{itemize}
\end{sloppypar}

%% file: tex/2related_work.tex
\section{Related work}

\noindent\textbf{Edge-cloud Collaboration.}
Deep learning applications are widely used~\cite{wang2017community,li2022end,li2023multi,li2023winner,wu2023focus,wu2023precedent,tang2024oodkd,qin2020health}, but they are fundamentally resource-intensive and difficult to deploy on the edge~\cite{tang2024modelgpt,chen2023learning,chen2023learning_arxiv,huang2022generspeech,huang2023make,DBLP:conf/ijcai/HuangL0S00Z22,cao2023_10.1145/3604237.3626868,li2023finetuning,li2022fine,lili_10.1145/3581783.3611847}, so edge-cloud collaboration~\cite{ref:edge_cloud_survey,zhangsyDBLP:conf/kdd/QianXLZJLZC022} is playing an increasingly important role. Cloud-based and on-edge machine learning are two distinct approaches with different benefits and drawbacks. Edge-cloud collaboration can take advantage of them and make them complement one another. Federated learning, such as FedAVG~\cite{ref:federated_fedavg}, is one of the most well-known forms of edge-cloud collaboration. Federated learning is also often used for various tasks such as multi-task learning~\cite{ref:federated_multi_task,ref:federated_multi_task2}, etc. But the federated learning method for edge-cloud collaboration is too rigid for many real-world scenarios. \cite{ref:edge_cloud} designs multiple models with the same functions but different training processes, and a Meta Controller is used to determine which model should be used. EC-CDR, such as DUET~\cite{ref:duet}, draw inspiration from the HyperNetwork concept, ensuring that models on the edge can generalize well to the current data distribution at every moment without the need for any training on the edge. However, high request frequency and low communication revenue significantly reduce their practicality. This paper focuses on addressing these shortcomings of EC-CDR.

\noindent\textbf{Sequential Recommendation.}
Sequential recommendation models the user's historical behavior sequence. 
Previous sequential recommendation algorithm such as ~\cite{ref:fpmc} and ~\cite{latifi2021session} are non-deep learning based and uses Markov decision chains to model behavioral sequences. To improve the performance of the model, recent works~\cite{ref:gru4rec,ref:din,ref:sasrec,ref:bert4rec,ref:srgnn,ref:surge,zhangsy2023personalized,zhangsyDBLP:conf/mm/ZhangJWKZZYYW20,chen2021deep,lv2023parameters,lv2022personalizing,su2023enhancing,su2023personalized,ji2023online,ji2023partial,li2023propensity,li2024kernel,lin2023mitigating,lin2023temporally} propose the sequential recommendation model based on deep learning. Among them, the most well-known sequence recommendation models are as follows: GRU4Rec~\cite{ref:gru4rec} uses GRU to model behavior sequences and achieves excellent performance. DIN~\cite{ref:din} and SASRec~\cite{ref:sasrec} algorithms, respectively, introduce attention and transformer into sequential recommendation, which is fast and efficient. These methods are relatively influential in both academia and industry. In practical settings, deploying recommendation models at the edge faces constraints due to limited parameters and complexity, alongside the need for real-time operation which hampers real-time model updates using conventional methods. This impacts the model's generalization capability across different data distributions. This paper explores methods to lower communication costs for a more efficient EC-CDR paradigm.

%% file: tex/3method.tex
\section{Methodology}
We describe the proposed \method{} in this section by presenting each module and introduce the learning strategy of \method{}.

\begin{figure*}[!h]
  \centering
\includegraphics[width=0.88\linewidth]{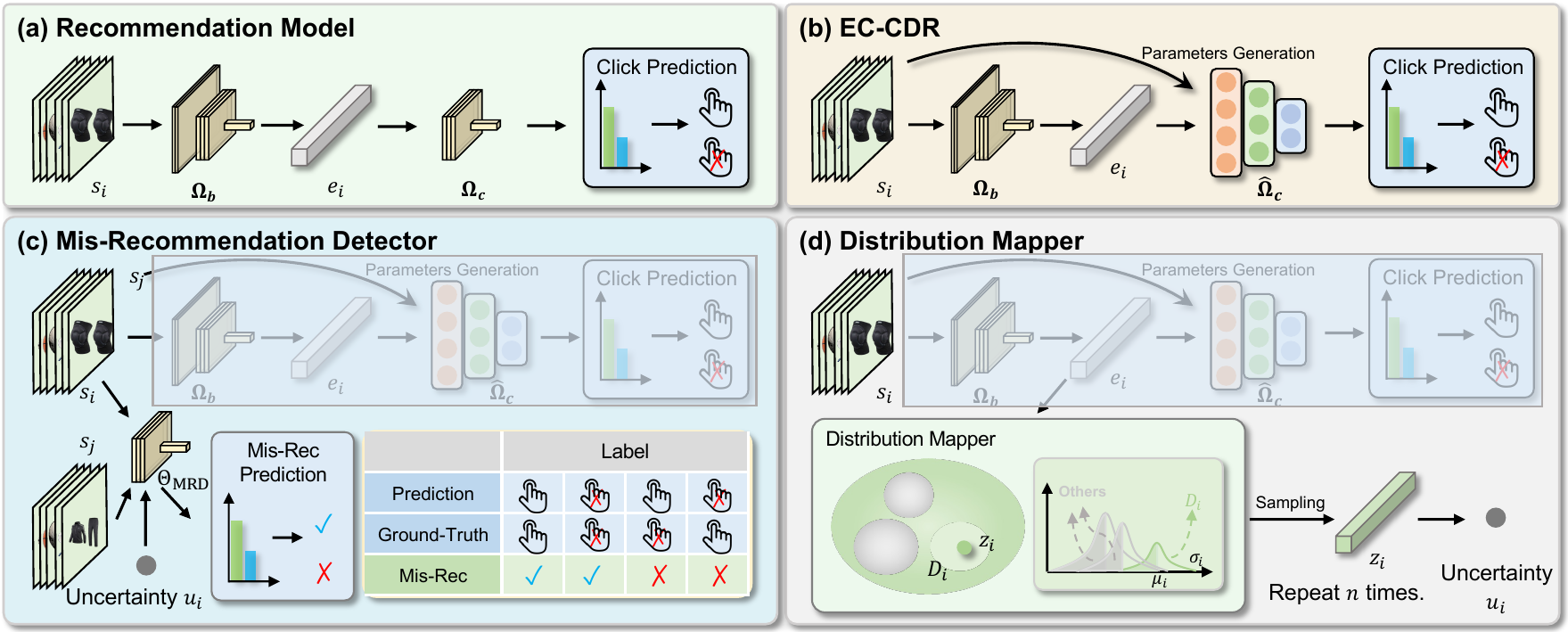}
\vspace{-0.35cm}
  \caption{Overview of the \textbf{IntellectReq}. (a) describes the conventional recommendation model. (b) describes the EC-CDR. (c) and (d) respectively showcase the IntellectReq, and its Mis-Recommendation Detector and Distribution Mapper modules.
\vspace{-0.2cm}
  }
    \label{fig:architecture_compare}
\vspace{-0.3cm}
\end{figure*}

\subsection{Problem Formulation}

\begin{sloppypar}
In EC-CDR, we have access to a set of edges $\mathcal{D}=\{d^{(i)}\}_{i=1}^{\mathcal{N}_d}$, where each edge with its personal i.i.d history samples  $\mathcal{S}_{H^{(i)}}=\{x^{(j,t)}_{H^{(i)}}=\{u^{(j)}_{H^{(i)}},v^{(j)}_{H^{(i)}},s^{(j,t)}_{H^{(i)}}\}, y^{(j)}_{H^{(i)}}\}_{j=1}^{\mathcal{N}_{H^{(i)}}}$ and real-time samples $\mathcal{S}_{R^{(i)}}=\{x^{(j,t)}_{R^{(i)}}=\{u^{(j)}_{R^{(i)}},v^{(j)}_{R^{(i)}},s^{(j,t)}_{R^{(i)}}\}\}_{j=1}^{\mathcal{N}_{R^{(i)}}}$ in the current session, where $\mathcal{N}_{d}$, $\mathcal{N}_{H^{(i)}}$ and $\mathcal{N}_{R^{(i)}}$ represent the number of edges, history data, and real-time data, respectively. $u$, $v$ and $s$ represent user, item and click sequence composed of items. It should be noted that $s^{(j,t)}$ represents the click sequence at moment $t$ in the $j$-th sample.The goal of EC-CDR is to generalize a trained global cloud model $\mathcal{M}_g(\cdot;\Theta_g)$ learned from $\{\mathcal{S}_{H^{(i)}}\}_{i=1}^{\mathcal{N}_d}$ to each specific local edge model $\mathcal{M}_{d^{(i)}}(\cdot;\Theta_{d^{(i)}})$ conditioned on real-time samples $\mathcal{S}_{R^{(i)}}$, where $\Theta_g$ and $\Theta_{d^{(i)}}$ respectively denote the learned parameters for the global cloud model and local edge model. 
\end{sloppypar}
\begin{equation}
\resizebox{0.41\textwidth}{!}{
$
\textbf{EC-CDR}:
\underbrace{\mathcal{M}_{g}(\{\mathcal{S}_{H^{(i)}}\}_{i=1}^{\mathcal{N}_d};\Theta_g)}_{\rm{Global\ Cloud\ Model}}  
\stackrel[\text{Parameters}]{\text{Data}}{\longleftrightarrows}
\underbrace{\mathcal{M}_{d^{(i)}}(\mathcal{S}_{R^{(i)}};\Theta_{d^{(i)}})}_{\rm{Local\ Edge\ Model}}.
$
}
 \label{eq:problem_formulation_duet}
\end{equation}
To determine whether to request parameters from the cloud, 
IntellectReq uses $\mathcal{S}_{_{\rm{MRD}}}$ to learn a Mis-Recommendation Detector, which decides whether to update the edge model by the EC-CDR framework. $\mathcal{S}_{_{\rm{MRD}}}$ is the dataset constructed based on $\mathcal{S}_{H}$ without any additional annotations for training IntellectReq. $\Theta_{_{\rm{MRD}}}$ denotes the learned parameters for the local MRD model.
\begin{equation}
\resizebox{0.435\textwidth}{!}{
$
\rm {\textbf{IntellectReq}}:
 \underbrace{\mathcal{M}_{c^{(i)}_{t}}(\mathcal{S}_{_{\rm{MRD}}};\Theta_{_{\rm{MRD}}})}_{\rm{Local\ Edge\ Model}} \xrightarrow{\text{Control}} \underbrace{(\mathcal{M}_{g} 
\stackrel[\text{Parameters}]{\text{Data}}{\longleftrightarrows}
 \mathcal{M}_{d^{(i)}})}_{\rm{EC-CDR}}.
 $
 }
 \label{eq:problem_formulation_mrd}
\end{equation}

\subsection{IntellectReq}
Figure~\ref{fig:architecture_compare} is the overview of Recommendation model, EC-CDR, and IntellectReq framework which consists of Mis-Recommendation Detector (MRD) and Distribution Mapper (DM) to achieve high revenue under any requested budget.
We first introduce the EC-CDR, and then present IntellectReq, which we propose to overcome the frequent and low-revenue drawbacks of EC-CDR requests. IntellectReq achieves high communication revenue under any edge-cloud communication budget in EC-CDR. MRD can determine whether to request parameters from the cloud model $\mathcal{M}_g$ or to use the edge recommendation model $\mathcal{M}_d$ based on real-time data $\mathcal{S}_{R^{(i)}}$. DM helps MRD make further judgments by discriminating the uncertainty in the recommendation model's understanding of data semantics.

\label{subsec:ms}
\subsubsection{The framework of EC-CDR}
\label{subsec:dc_cdr}
In EC-CDR, a recommendation model with a static layers and a dynamic layers will be trained for the global cloud model development. The goal of the EC-CDR can thus be formulated as the following optimization problem:
\begin{equation}
\begin{aligned}
\centering
& 
\resizebox{0.22\textwidth}{!}{
$
\hat{y}^{(j)}_{H^{(i)}} = f_{\rm rec}(\Omega(x^{(j)}_{H^{(i)}};\Theta_g^b); \Theta_g^c), 
$}
\\ 
& 
\resizebox{0.27\textwidth}{!}{
$
\mathcal{L}_{\rm{rec}}=\sum_{i=1}^{\mathcal{N}_d} \sum_{j=1}^{\mathcal{N}_{R^{(i)}}} 
D_{ce} (y^{(j)}_{H^{(i)}}, \hat{y}^{(j)}_{H^{(i)}}),
$}
\label{eq:umn_loss}
\end{aligned}
\end{equation}
where $D_{ce}(\cdot;\Theta_g^b)$ denotes the cross-entropy between two probability distributions, $f_{\rm rec}(\cdot)$ denotes the dynamic layers of the recommendation model, $\Omega(x^{(j)}_{H^{(i)}};\Theta_g^b)$ is the static layers extracting features from $x^{(j)}_{H^{(i)}}$. EC-CDR is decoupled edge-model with a ``static layers'' and ``dynamic layers'' training scheme to achieve better personalization. 
The primary factor enhancing the on-edge model's generalization to real-time data through EC-CDR is its dynamic layers. Upon completion of training, the static layers' parameters remain static, denoted as $\Theta_g^b$, as determined by Eq.~\ref{eq:umn_loss}. Conversely, the dynamic layers' parameters, represented by $\Theta_g^c$, are dynamically generated based on real-time data by the cloud generator.

\textit{In edge inference}, the cloud-based parameter generator uses the real-time click sequence $s^{(j,t)}_{R^{(i)}} \in \mathcal{S}_{R^{(i)}}$ to generate the parameters,
\begin{align}
\label{eq:lightweight_encoder}
    \boldsymbol{h}^{(n)}_{R^{(i)}} = {L}^{(n)}_{\rm{layer}}(\boldsymbol{e}^{(j,t)}_{R^{(i)}} = {E}_{\rm{shared}} (s^{(j,t)}_{R^{(i)}})), \forall n=  1, \cdots, \mathcal{N}_l,
\end{align}
where ${E}_{\rm{share}}(\cdot)$ represents the shared encoder. ${L}^{(n)}_{\rm{layer}}(\cdot)$ is a linear layer used to adjust $\boldsymbol{e}^{(j,t)}_{R^{(i)}}$ which is the output of ${E}_{\rm{share}}(\cdot)$ to the $n^{th}$ dynamic layer features. $\boldsymbol{e}^{(j,t)}_{R^{(i)}}$ means embedding vector generated by the click sequence at the moment $t$. 

The cloud generator model treats the parameters of a fully-connected layer as a matrix $K^{(n)} \in \mathbb{R}^{N_{in}\times N_{out}}$, where $N_{in}$ and $N_{out}$ represent the number of input neurons and output neurons of the $n^{th}$ fully-connected layers, respectively. 
Then the cloud generator model $g(\cdot)$ converts the real-time click sequence $s^{(j,t)}_{R^{(i)}}$ into dynamic layers parameters $\hat{\Theta}_g^c$ by $K^{(n)}_{R^{(i)}} = g^{(n)}(\boldsymbol{e}^{(n)}_{R^{(i)}})$. Since the following content no longer needs the superscript $(n)$, we simplify $g(\cdot)$ to $g(\cdot)={L}^{(n)}_{\rm{layer}}({E}_{\rm{shared}}(\cdot))$. Then, the edge recommendation model updates the parameters and makes inference as follows,
\begin{align}
\label{eq:ppg_pred}
\hat{y}^{(j,t)}_{R^{(i)}}=f_{\rm rec}(\Omega(x^{(j,t)}_{R^{(i)}};\Theta_g^b); \hat{\Theta}_g^c=g(s^{(j,t)}_{R^{(i)}};\Theta_p)).
\end{align}

\begin{figure*}[!h]
  \centering
\includegraphics[width=0.88\linewidth]{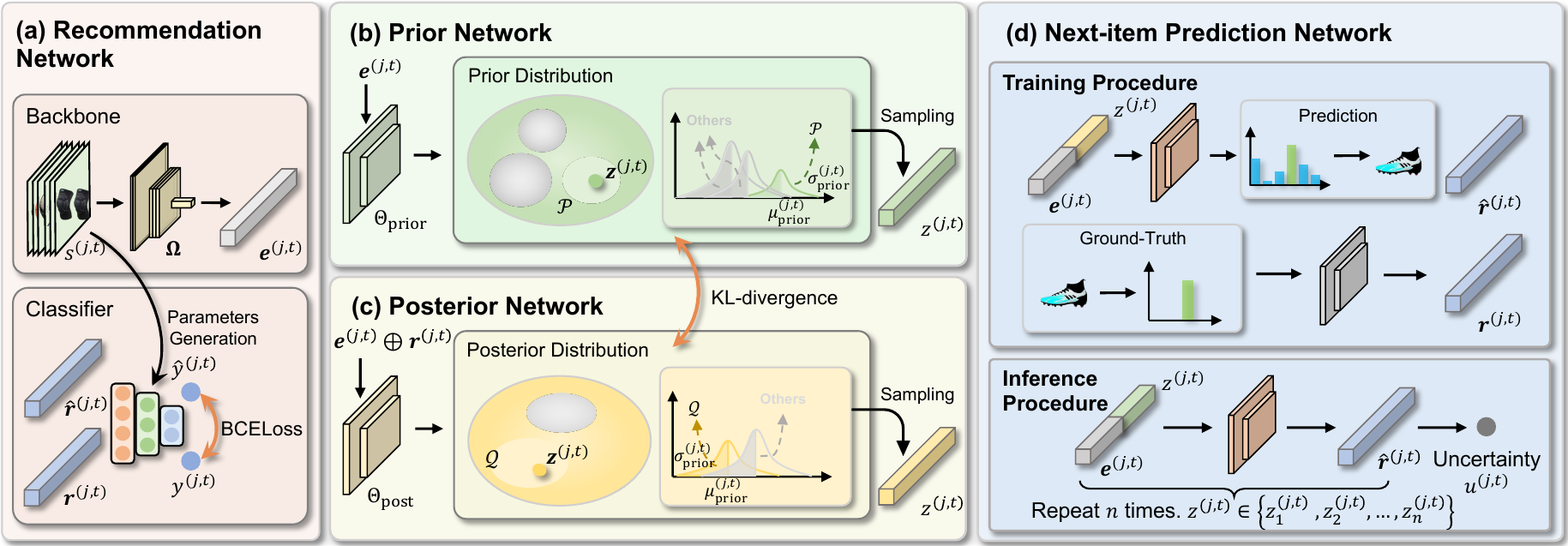}
  \vspace{-0.3cm}
  \caption{Overview of the proposed Distribution Mapper.
  \textbf{Training procedure:} The architecture includes Recommendation Network, Prior Network, Posterior network and Next-item Perdition Network. Loss consists of the classification loss and the KL-Divergence loss. \textbf{Inference procedure:} The architecture includes Recommendation Network, Prior Network and Next-item Perdition Network. The uncertainty is calculated by the multi-sampling output.
  }
  \label{fig:uncertainty_architecture}
  \vspace{-0.3cm}
\end{figure*}

\textit{In cloud training}, all layers of the cloud generator model are optimized together with the static layers of the primary model  that are conditioned on the global history data
$\mathcal{S}_{H^{(i)}}=\{x^{(j)}_{H^{(i)}}, y^{(j)}_{H^{(i)}}\}_{j=1}^{\mathcal{N}_{H^{(i)}}}$,  instead of optimizing the static layers of the primary model first and then optimizing the cloud generator model. 
The cloud generator model loss function
is defined as follows:
\begin{align}
    \label{eq:loss_func}
\resizebox{0.26\textwidth}{!}{
$
\mathcal{L}=\sum_{i=1}^{\mathcal{N}_d} \sum_{j=1}^{\mathcal{N}_{H^{(i)}}}
 D_{ce} (y^{(j)}_{H^{(i)}}, \hat{y}^{(j)}_{H^{(i)}}).
 $}
\end{align}

EC-CDR could improve the generalization ability of the edge recommendation model.
However, EC-CDR could not be easily deployed in a real-world environment due to the high request frequency and low communication revenue. Under the EC-CDR framework, the moment $t$ in Eq.~\ref{eq:ppg_pred} is equal to the current moment $T$, which means that the edge and the cloud communicate at every moment. 
In fact, however, a lot of communication is unnecessary because $\hat{\Theta}_g^c$ generated by the sequence earlier may work well enough.
To alleviate this issue, we propose MRD and DM to solve the problem when the edge recommendation model should update parameters.

\subsubsection{Mis-Recommendation Detector}
\label{subsec:mrd}
\begin{sloppypar}
The training procedure of MRD can be divided into two stages.
The goal of the first stage is to construct a MRD dataset $\mathcal{S}_C$ based on the user's historical data without any additional annotation to train the MRD. 
The cloud model $\mathcal{M}_{g}$ and the edge model $\mathcal{M}_{d}$ are trained in the same way as the training procedure of EC-CDR. 
\end{sloppypar}
\begin{align}
\label{eq:mr_dataset1}
\resizebox{0.32\textwidth}{!}{
$
\hat{y}^{(j,t,t')}_{H^{(i)}}=f_{\rm{rec}}(\Omega(x^{(j,t)}_{H^{(i)}};\Theta_g^b); \hat{\Theta}_g^c=g(s^{(j,t')}_{H^{(i)}};\Theta_p)).
$}
\end{align}
Here, we set $t'\leq t=T$. That is, when generating model parameters, we use the click sequence $s^{(j,t')}_{R^{(i)}}$ at the previous moment $t'$, but this model is used to predict the current data. Then we can get $c^{(j,t,t')}$ that means whether the sample be correctly predicted based on the prediction $\hat{y}^{(j,t,t')}_{R^{(i)}}$ and the ground-truth $y^{(j,t)}_{R^{(i)}}$.
\begin{equation}
\begin{aligned}
\resizebox{0.24\textwidth}{!}{
$
    c^{(j,t,t')}=\left\{
        \begin{array}{lll}
        1, \hat{y}^{(j,t,t')}_{R^{(i)}}=y^{(j,t)}_{R^{(i)}}; \\
        0, \hat{y}^{(j,t,t')}_{R^{(i)}} \neq y^{(j,t)}_{R^{(i)}}.
        \end{array}
    \right.
    ,
$
}
\label{eq:mr_dataset2}
\end{aligned}
\end{equation}
\begin{equation}
\begin{aligned}
\resizebox{0.4\textwidth}{!}{
$
\mathcal{L}_{_{\rm{MRD}}}= \sum_{j=1}^{{|\mathcal{S}_{_{\rm{MRD}}}^{(i)}}|}
    \sum_{t'=1}^{T}
    l(y_j, \hat{y}=f_{_{\rm{MRD}}}(s^{(j,t)}, s^{(j,t')};\Theta_{_{\rm{MRD}}})).
$}
 \label{eq:mr_train}
\end{aligned}
\end{equation}
Then we construct the new mis-recommendation training dataset as follows:
$\mathcal{S}_{_{\rm{MRD}}}^{(i)}=\{s^{(j,t)}, s^{(j,t')}, c^{(j,t,t')}\}_{0\leq t'\leq t=T}.$ 
Then, a dynamic layers $f_{_{\rm{MRD}}}(\cdot)$ can be trained on $\mathcal{S}_{_{\rm{MRD}}}^{(i)}$ according to the Eq.~\ref{eq:mr_train}, where $t=T$ and the loss function $l(\cdot)$ is cross entropy.

\subsubsection{Distribution Mapper}
\label{subsec:dm}
Although the MRD could determine when to update edge parameters, it is insufficient to simply map a click sequence to a certain representation in a high-dimensional space due to ubiquitous noises in click sequences. So we design the DM as Figure~\ref{fig:uncertainty_architecture} make it possible to directly perceive the data distribution shift and determine the uncertainty in the recommendation model's understanding of the semantics of the data. 

Inspired by Conditional-VAE, we map click sequences to normal distributions. Different from the MRD, the DM module consider a variable {$u^{(j,t)}$} to denote the uncertainty in Eq.~\ref{eq:mr_train} as:
\begin{equation}
\begin{aligned}
\resizebox{0.43\textwidth}{!}{
$
    \mathcal{L}_{_{\rm{MRD}}}= \sum_{j=1}^{{|\mathcal{S}_{_{\rm{MRD}}}^{(i)}}|}
    \sum_{t'=1}^{T}
    l(y_j, \hat{y}=f_{_{\rm{MRD}}}(s^{(j,t)}, s^{(j,t')},u^{(j,t)};\Theta_{_{\rm{MRD}}})).
$}
 \label{eq:mru_train}
\end{aligned}
\end{equation}
The uncertainty variable \textbf{$u^{(j,t)}$} shows the recommendation model's understanding of the semantics of the data. DM focuses on how to learn such uncertainty variable {$u^{(j,t)}$}.

Distribution Mapper consists of three components as shown in the figure in Appendix, namely the \underline{Pr}ior \underline{N}etwork $P(\cdot)$ (PRN), the \underline{Po}sterior \underline{N}etwork $Q(\cdot)$ (PON), and the \underline{N}ext-item \underline{P}rediction \underline{N}etwork $f(\cdot)$ (NPN) that includes the static layers $\Omega(\cdot)$ and dynamic layers $f_{_{\rm{NPN}}}(\cdot)$. Note that $\Omega(\cdot)$ here is the same as $\Omega(\cdot)$ in section~\ref{subsec:dc_cdr} and \ref{subsec:mrd}, so there is almost no additional resource consumption. We will first introduce the three components separately, and then introduce the training procedure and inference procedure. 

\noindent \textit{Prior Network.}
The Prior Network with weights $\Theta_{\rm{prior}}$ and $\Theta^{'}_{\rm{prior}}$ maps the representation of a click sequence $s^{(j,t)}$ to a prior probability distribution. We set this prior probability distribution as a normal distribution with mean $\mu_{\rm prior}^{(j,t)}=\Omega_{\rm{prior}}(s^{(j,t)};\Theta_{\rm prior})\in \mathbb{R}^N$ and variance $\sigma_{\rm prior}^{(j,t)}=\Omega^{'}_{\rm{prior}}(s^{(j,t)};\Theta^{'}_{\rm prior})\in \mathbb{R}^N$.
\begin{equation}
\resizebox{0.26\textwidth}{!}{
$
\mathbf{z}^{(j,t)} \sim P(\cdot|s^{(j,t)})=\mathcal{N}(\mu_{\rm{prior}}^{(j,t)},\sigma_{\rm{prior}}^{(j,t)}).
$}
\label{eq:uncertainty_prior}
\end{equation}

\begin{sloppypar}
\noindent \textit{Posterior Network.}
The Posterior Network $\Omega_{\rm{post}}$ with weights $\Theta_{\rm{post}}$ and $\Theta^{'}_{\rm{post}}$ can enhance the training of the Prior Network by introducing posterior information. It maps the representation concatenated by the representation of the next-item $r^{(j,t)}$ and of the click sequence $s^{(j,t)}$ to a normal distribution. 
we define the posterior probability distribution as a normal distribution with mean $\mu_{\rm post}^{(j,t)}=\Omega_{\rm{post}}(s^{(j,t)};\Theta_{\rm post})\in \mathbb{R}^N$ and variance $\sigma_{\rm post}^{(j,t)}=\Omega^{'}_{\rm{post}}(s^{(j,t)};\Theta^{'}_{\rm post})\in \mathbb{R}^N$. 
\end{sloppypar}

\begin{equation}
 \begin{aligned}
 \resizebox{0.28\textwidth}{!}{
$
\mathbf{z}^{(j,t)} \sim Q(\cdot|s^{(j,t)},r^{(j,t)})=\mathcal{N}(\mu_{\rm post}^{(j,t)},\sigma_{\rm post}^{(j,t)}).
\label{eq:uncertainty_post}
$}
\end{aligned}
\end{equation}
\noindent \textit{Next-item Prediction Network.}
The Next-item Prediction Network with weights $\Theta_c$ predicts the embedding of the next item $\hat{r}^{(j,t)}$ to be clicked based on the user's click sequence $s^{(j,t)}$ as follows,
\begin{equation}
 \begin{aligned}
&~~~~~~~~~\hat{r}^{(j,t)}=f_c(\boldsymbol{e}^{(j,t)}=\Omega(s^{(j,t)};\Theta_b), z^{(j,t)};\Theta_c), \\
&\hat{y}^{(j,t)} = f_{\rm rec}(\Omega(x^{(j,t)};\Theta_g^b), \hat{r}^{(j,t)};g(\boldsymbol{e}^{(j,t)};\Theta_p)).
\label{eq:uncertainty_item_pred}
\end{aligned}
\end{equation}

\noindent \textbf{Training Procedure.}
In the training procedure, two losses need to be constructed, one is recommendation prediction loss $\mathcal{L}_{rec}$ and the other is distribution difference loss $\mathcal{L}_{dist}$.
Like the way that most recommendation models are trained, $\mathcal{L}_{rec}$ uses the binary cross-entropy loss function $l(\cdot)$ to penalize the difference between $\hat{y}^{(j,t)}$ and $y^{(j,t)}$. The difference is that here NPN uses the feature $z$ sampled from the prior distribution $Q$ to replace $e$ in formula 5
In addition, $\mathcal{L}_{dist}$ penalizes the difference between the posterior distribution $Q$ and the prior distribution $P$ with the help of the Kullback-Leibler divergence.
$\mathcal{L}_{dist}$ "pulls" the posterior and prior distributions towards each other. The formulas for $\mathcal{L}_{rec}$ and $\mathcal{L}_{dist}$ are as follows,
\begin{equation}
 \begin{aligned}
\mathcal{L}_{rec}=\mathbb{E}_{\mathbf{z} \sim Q(\cdot|s^{(j,t)},y^{(j,t)})}[l(y^{(j,t)}|\hat{y}^{(j,t)})],
\label{eq:uncertainty_loss_1}
\end{aligned}
\end{equation}
\begin{equation}
 \begin{aligned}
\mathcal{L}_{dist}=D_{KL}(Q(z|s^{(j,t)},y^{(j,t)})||P(z|s^{(j,t)})).
\label{eq:uncertainty_loss_2}
\end{aligned}
\end{equation}
Finally, we optimize DM according to,
\begin{equation}
 \begin{aligned}
\mathcal{L}(y^{(j,t)},s^{(j,t)})=\mathcal{L}_{rec}+\beta \cdot \mathcal{L}_{dist}.
\label{eq:uncertainty_loss}
\end{aligned}
\end{equation}
During training, the weights are randomly initialized.

\noindent \textbf{Inference Procedure}. In the inference procedure, the posterior network will be removed from DM because there is no posterior information during the inference procedure. Uncertainty variable $u^{(j,t)}$ is calculated by the multi-sampling outputs as follows:
\begin{equation}
\begin{aligned}
u^{(j,t)}={\rm var}(\hat{r}_i=f_c(\Omega(s^{(j,t)};\Theta_b), z^{(j,t)}_{1\sim n};\Theta_c)), 
\label{eq:uncertainty_inference}
\end{aligned}
\end{equation}
where $n$ denotes the sampling times. Specifically, we consider the dimension of $\hat{r}^{(j,t)}$ is $N\times 1$, $\hat{r}_{i}^{(j,t),(k)}$ as the $k$-th value of the $\hat{r}_{i}^{(j,t)}$ vector, and calculate the variance as follows:
\begin{equation}
\begin{aligned}
\resizebox{0.215\textwidth}{!}{
$
{\rm var}(\hat{r}_i)=\sum_{k=1}^{N}{\rm var}\hat{r}_{1\sim n}^{(j,t),(k)} .
$}
\label{eq:uncertainty_inference_detail}
\end{aligned}
\end{equation}

\subsubsection{On-edge Model Update}
\textbf{M}is-\textbf{R}ecommendation \textbf{S}core~(MRS) is a variable calculated based on the output of MRD and DM, which directly affects whether the model needs to be updated.
\begin{equation}
\begin{aligned}
\resizebox{0.55\linewidth}{!}{
$
   {\rm{MRS}}=1-f_{{\rm{MRD}}}(s^{(j,t)}, s^{(j,t')};\Theta_{_{\rm{MRD}}})
$
}
 \label{eq:mrs_}
\end{aligned}
\end{equation}
\begin{equation}
\begin{aligned}
\resizebox{0.45\linewidth}{!}{
$
   \rm{Update}=\mathds{1}(MRS \leq Threshold)
$}
 \label{eq:update_mrs}
\end{aligned}
\end{equation}
In the equation above, $\mathds{1}(\cdot)$ is the indicator function. 
To get the threshold, we need to collect user data for a period of time, then get the MRS values corresponding to these data on the cloud and sort them, and then set the threshold according to the load of the cloud server. For example, if the load of the cloud server needs to be reduced by 90\%, that is, when the load is only 10\% of the previous value, only the minimum 10\% position value needs to be sent to each edge as the threshold. During inference, each edge determines whether it needs to update the edge model based on equation~\ref{eq:mrs_} and \ref{eq:update_mrs}, that is, whether it needs to request new parameters.

%% file: tex/4experiments.tex
\section{Experiments}

We conducted extensive experiments to evaluate the effectiveness and generalizability of the proposed 
IntellectReq. We put part of the experimental setup, results and analysis in the Appendix.
\subsection{Experimental Setup.}
\begin{sloppypar}
\noindent\textbf{Datasets.} We evaluate on \texttt{Amazon CDs~(CDs)}, \texttt{Amazon Electronic~(Electronic)}, \texttt{Douban Book~(Book)},
three widely used public benchmarks in the recommendation tasks. 
\end{sloppypar}
\noindent \textbf{Evaluation Metrics} 
In the experiments, we use the widely adopted AUC~\footnote{Note 0.1\% absolute AUC gain is regarded as significant for the CTR task~\cite{ref:apg_rs1,ref:duet,ref:sasrec,ref:din}\label{fn:auc}}, UAUC$^{~\ref{fn:auc}}$, HitRate and NDCG as the metrics. 
    
\noindent\textbf{Baselines.}
To verify the applicability, the following representative sequential modeling approaches are implemented and compared with the counterparts combined with the proposed method.
    \
\newline
    \textbf{DUET}~\cite{ref:duet} and \textbf{APG}~\cite{ref:apg_rs1} are SOTA of EC-CDR, which generate parameters through the edge-cloud collaboration for different tasks. With the cloud generator model, the on-edge model could generalize well to the current data distribution in each session without training on the edge. \textbf{GRU4Rec}~\cite{ref:gru4rec}, \textbf{DIN}~\cite{ref:din}, and \textbf{SASRec}~\cite{ref:sasrec} are three of the most widely used sequential recommendation methods in the academia and industry, which respectively introduce GRU, Attention, and Self-Attention into the recommendation system. \textbf{LOF}~\cite{ref:lof} and \textbf{OC-SVM}~\cite{ref:ocsvm} estimate the density of a given point via the ratio of the local reachability of its neighbors and itself. They can be used to detect changes in the distribution of click sequences. \\ 
For the IntellectReq, we consider SASRec as edge-model unless otherwise stated, but note that IntellectReq broadly applies to lots of sequential recommendation model such as DIN, GRU4Rec, etc.

\noindent\textbf{Evaluation Metrics.}
    We use the widely adopted AUC, HitRate, and NDCG as the metrics to evaluate model performance. 

\subsection{Experimental Results.}

\subsubsection{Quantitative Results.}
\label{subsubsec:quantitative_results}
\begin{figure*}[!h]
  \centering
\includegraphics[width=\linewidth]{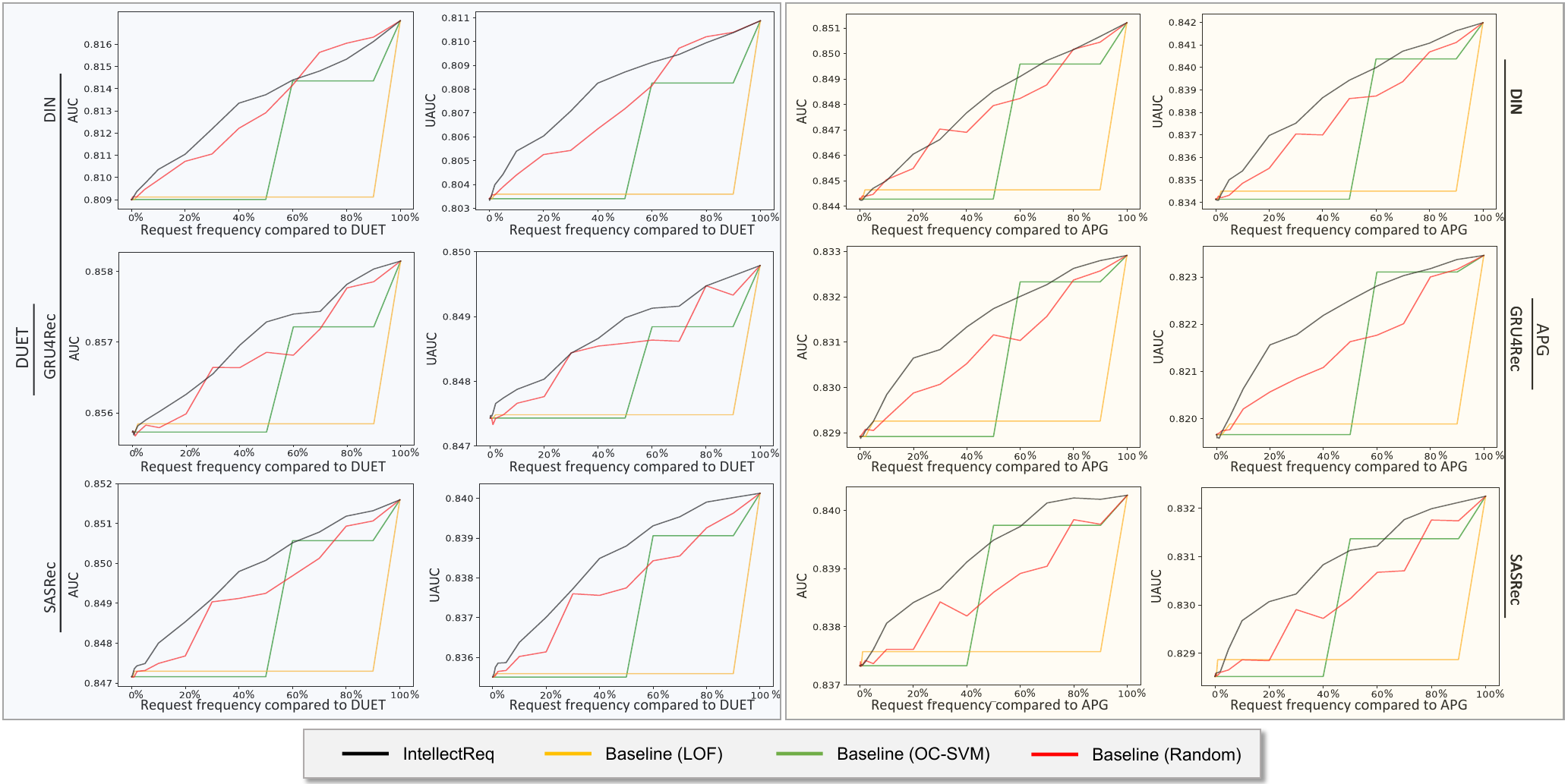}
\vspace{-0.7cm}
  \caption{Performance \textit{w.r.t.} Request Frequency curve based on previous 1 time difference on-edge dynamic model.}
  \label{fig:request_curve_duet}
  \vspace{-0.3cm}
\end{figure*}

\begin{figure}[!h]
  \centering
\includegraphics[width=0.94\linewidth]{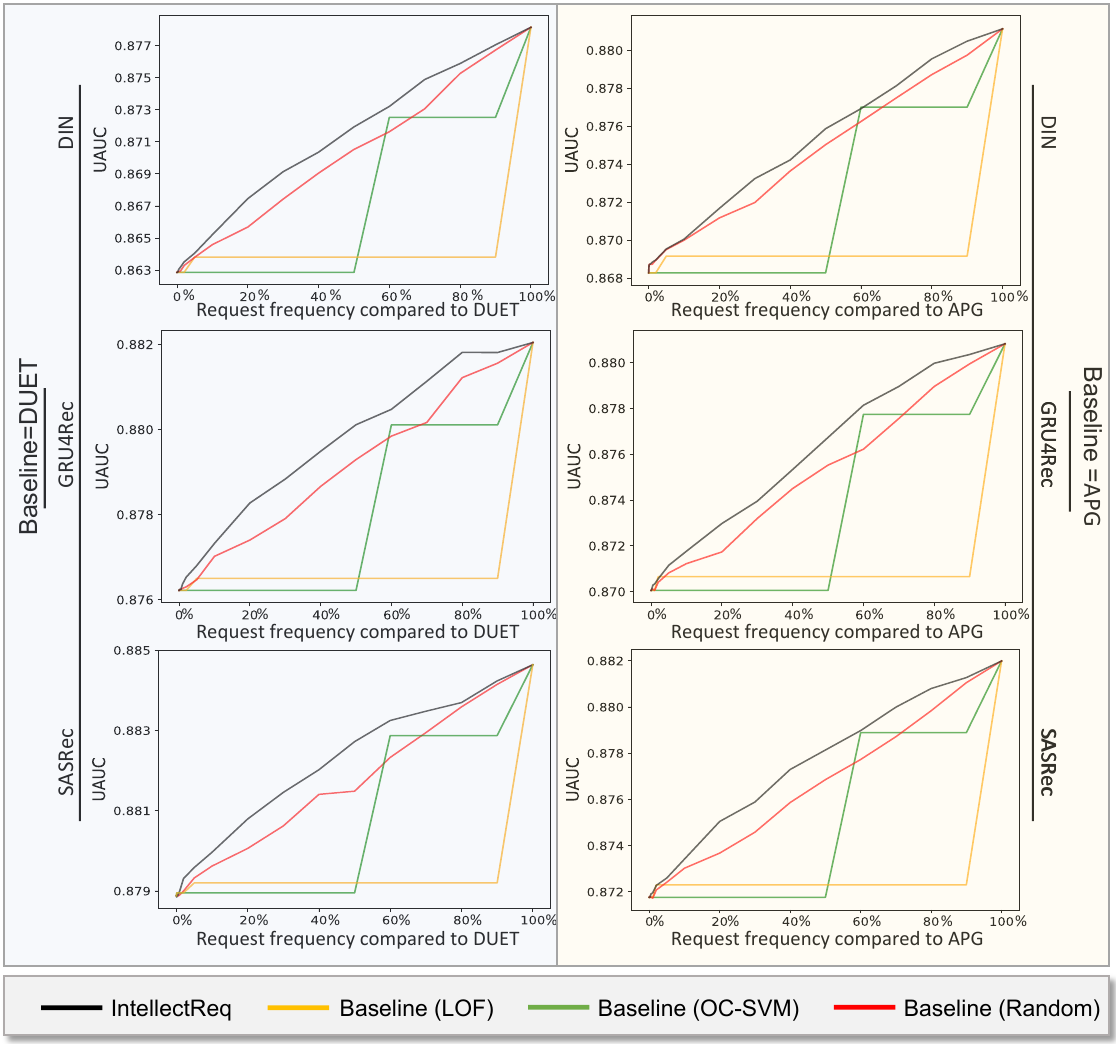}
\vspace{-0.3cm}
  \caption{Performance \textit{w.r.t.} Request Frequency based on previous 1 time difference on-edge dynamic model.}
  \label{fig:request_curve_apg}
\vspace{-0.6cm}
\end{figure}

\begin{figure}[!h]
  \centering
\includegraphics[width=0.94\linewidth]{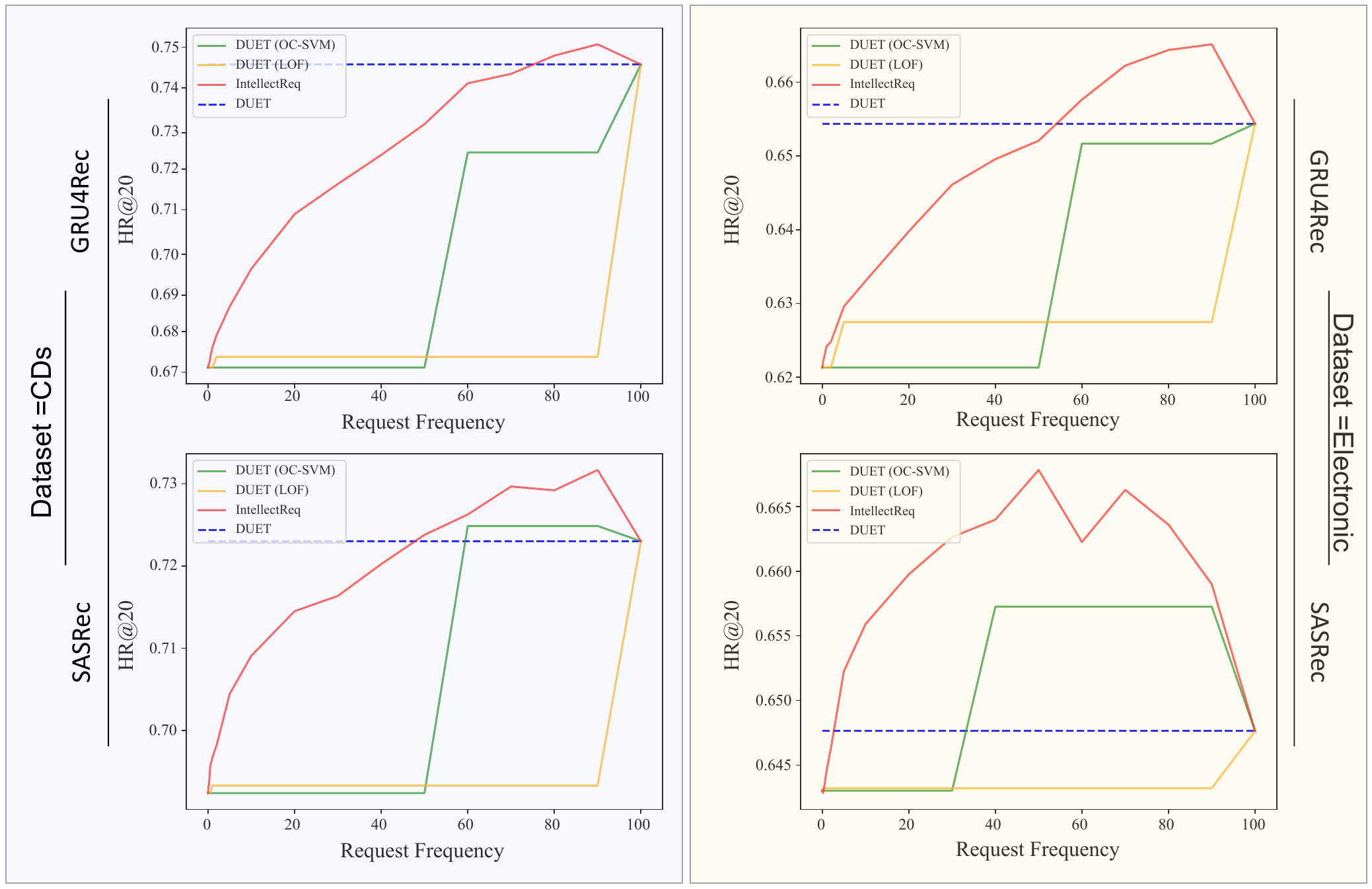}
\vspace{-0.3cm}
  \caption{Performance \textit{w.r.t.} Request Frequency based on on-edge static model.}
  \label{fig:request_curve_static}
\vspace{-0.2cm}
\end{figure}
Figure~\ref{fig:request_curve_duet}, \ref{fig:request_curve_apg}, and \ref{fig:request_curve_static} summarize the quantitative results of our framework and other methods on CDs and Electronic datasets. 
The experiments are based on state-of-the-art EC-CDR frameworks such as DUET and APG. As shown in Figure~\ref{fig:request_curve_duet}-\ref{fig:request_curve_apg}, we combine the parameter generation framework with three sequential recommendation models, DIN, GRU4Rec, SASRec. We evaluate these methods with AUC and UAUC metrics on CDs and Book datasets.
We have the following findings: 
(1) If all edge-model updated at $t-1$ moment, the DUET framework (DUET) and the APG framework (APG) can be viewed as the upper bound of performance for all methods since DUET and APG are evaluated with fixed 100\% request frequency and other methods are evaluated with increasing frequency. If all edge-model are the same as the cloud pretrained model, IntellectReq can even beat DUET, which indicates that in EC-CDR, not all edges need to be updated at every moment. In fact, model parameters generated by user data at some moments can be detrimental to performance.
Note that directly comparing the other methods with DUET and APG is not fair as DUET and APG use the fixed 100\% request frequency, which could not be deployed in lower request frequency.
(2) The random request method (DUET (Random), APG (Random)) works well with any request budget. However, it does not give the optimal request scheme for any request budget in most cases (such as Row.1). The correlation between its performance and Request Frequency tends to be linear. 
The performances of random request methods are unstable and unpredictable, where these methods outperform other methods in a few cases. 
(3) LOF (DUET (LOF), APG (LOF)) and OC-SVM (DUET (OC-SVM), APG (OC-SVM)) are two methods that could be used as simple baselines to make the optimal request scheme under a special and specific request budget.
However, they have two weaknesses. One is that they consume a lot of resources and thus significantly reduce the calculation speed. The other is they can only work under a specific request budget instead of an arbitrary request budget. For example, in the first line, the Request Frequency of OC-SVM can only be 
(4) In most cases, our IntellectReq can make the optimal request scheme under any request budget.

\subsubsection{Mis-recommendation score and profit.}
\label{sec:profit}

\begin{figure}
  \centering
\includegraphics[width=0.96\linewidth]{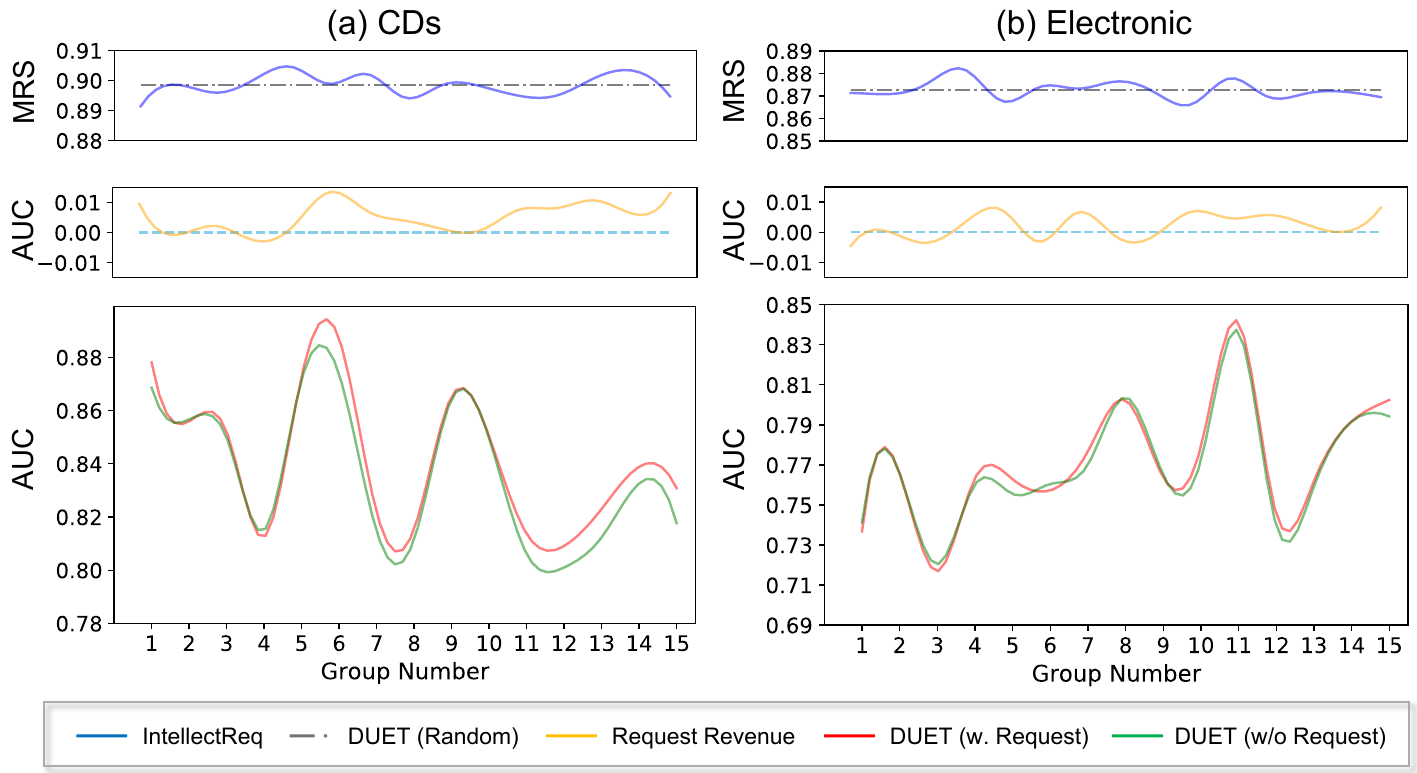}
\vspace{-0.3cm}
  \caption{Mis-Recommendation Score and Revenue.}
\label{fig:request_timing_and_earning}
\vspace{-0.3cm}
\end{figure}
To further study the effectiveness of MDR, we visualize the request timing and revenue in Figure~\ref{fig:request_timing_and_earning}.
As shown in Figure~\ref{fig:request_timing_and_earning}, we analyze the relationship between request and revenue. 
Every 100 users were assigned to one of 15 groups, which were selected at random. The Figure is divided into three parts, with the first part used to assess the request and the second and third parts used to assess the benefit.
The metric used here is Mis-Recommendation Score (MRS) to evaluate the request revenue. MRS is a metric to measure whether a recommendation will be made in error.
In other words, it can be viewed as an evaluation of the model's generalization ability. 
The probabilities of a mis-recommendation and requesting model parameters are higher and the score is lower.
\begin{sloppypar}
\begin{itemize}[itemsep=2pt,topsep=2pt,leftmargin=*]
    \item \textbf{IntellectReq} predicts the MRS based on the uncertainty and the click sequences at the moment $t$ and $t-1$.
    \item \textbf{DUET (Random)} randomly selects edges to request the cloud model to update the parameters of the edges. At this point, MRS can be considered as an arbitrary constant. We take the average value of IntellectReq's MRS as the MRS value.
    \item \textbf{DUET (w. Request)} represents all edge-model be updated at the moment $t$.
    \item \textbf{DUET (w/o. Request)} represents no edge-model be updated at moment $t-1$ in Figure~\ref{fig:request_curve_duet} and \ref{fig:request_curve_apg}, represents no edge-model be updated at moment $0$ in Figure~\ref{fig:request_curve_static}.
    \item \textbf{Request Revenue} represents the revenue, that is, DUET (w. Request) curve minus DUET (w/o Request).
\end{itemize} 
\end{sloppypar}

From Figure~\ref{fig:request_timing_and_earning}, we have the following observations:
(1) The trends of MRS and DUET Revenue are typically in the opposite direction, which means that when the MRS value is low, IntellectReq tends to believe that the edge's model cannot generalize well to the current data distribution. Then, the IntellectReq uses the most recent real-time data to request model parameters. As a result, the revenue at this time is frequently positive and relatively high. When the MRS value is high, IntellectReq tends to continue using the model that was updated at the previous moment $t-1$ instead of $t$ because it believes that the model on the edge can generalize well to the current data distribution. The revenue is frequently low and negative if the model parameters are requested at this point.
(2) Since the MRS of DUET (Random) is constant, it cannot predict the revenue of each request. The performance curve changes randomly because of the irregular arrangement order of groups.

\subsubsection{Ablation Study.}
\begin{figure}[!h]
\vspace{-0.3cm}
  \centering
\includegraphics[width=0.88\linewidth]{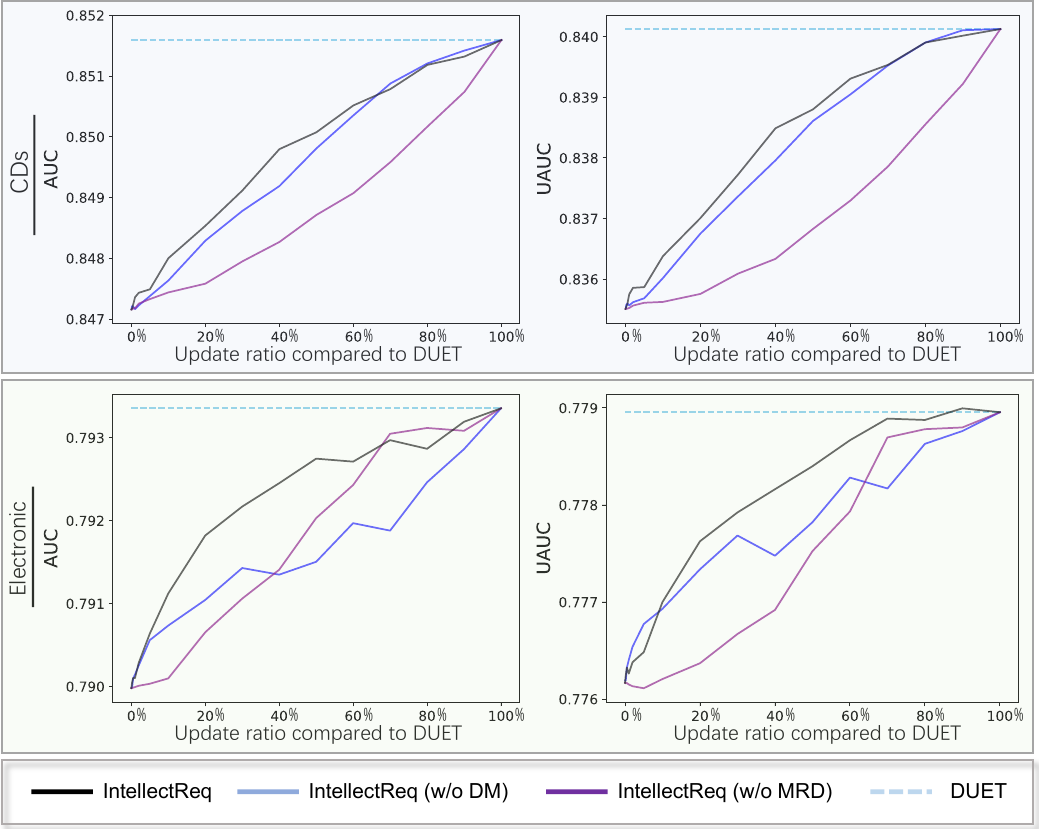}
\vspace{-0.35cm}
  \caption{Ablation study on model architecture.}
  \label{fig:ablation_study}
\vspace{-0.3cm}
\end{figure}
We conducted an ablation study to show the effectiveness of different components in IntellectReq. The results are shown in Figure~\ref{fig:ablation_study}.
We use \textbf{w/o.} and \textbf{w.} to denote without and with, respectively. From the table, we have the following findings: 
\begin{itemize}[itemsep=2pt,topsep=2pt,leftmargin=20pt]
\item \textbf{IntellectReq} means both DM and MRD are used.
\item \textbf{(\textbf{w/o.} DM)} means MRD is used but DM is not used.
\item \textbf{(\textbf{w/o.} MRD)} means DM is used but MRD is not used.
\end{itemize}
From the figure and table, we have the following observations: 
(1) Generally, IntellectReq achieves the best performance with different evaluation metrics in most cases, demonstrating the effectiveness of IntellectReq.
(2) When the request frequency is small, the difference between IntellectReq and IntellectReq (\textbf{w/o.} DM) is not immediately apparent, as shown in Fig.~\ref{fig:ablation_study}(d). The difference becomes more noticeable when the Request Frequency increases within a certain range. In brief, the difference exhibits the traits of first getting smaller, then larger, and finally smaller.

\subsubsection{Time and Space Cost.} Most edges have limited storage space, so the on-edge model must be small and sufficient.
The edge's computing power is rather limited, and the completion of the recommendation task on the edge requires lots of real-time processing, so the model deployed on the edge must be both simple and fast. Therefore, we analyze whether these methods are controllable and highly profitable based on the DUET framework, and additional time and space resource consumption under this framework is shown in Table~\ref{tab:time_space_cost}.
\begin{table}[!h]
\vspace{-0.2cm}
\caption{Extra Time and Space Cost on CDs dataset.}
\vspace{-0.35cm}
\label{tab:time_space_cost}
\resizebox{0.4\textwidth}{!}{
\begin{tabular}{c|c|c|c|c}
\toprule[2pt]
Method & Controllable & Profitable & Time Cost & Space Cost (Param.) \\
\midrule[1pt]
\midrule[1pt]
LOF & \XSolidBrush & \Checkmark & $225\rm{s}/11.3\rm{ms}$ &  $\approx0$ \\ \hline
OC-SVM & \XSolidBrush & \Checkmark & $160\rm{s}/9.7\rm{ms}$ & $\approx0$ \\ \hline
Random & \Checkmark & \XSolidBrush & $0\rm{s}/0.8\rm{ms}$ & $\approx0$ \\ \hline
\rowcolor[HTML]{F8F8F8} IntellectReq & \Checkmark & \Checkmark & $11\rm{s}/7.9\rm{ms}$ & $\approx 5.06k$ \\
\bottomrule[2pt]
\end{tabular}
}
\vspace{-0.3cm}
\end{table}
In the time consumption column, signal ``/'' separates the time consumption of cloud preprocessing and edge inference. Cloud preprocessing means that the cloud server first calculates the MRS value based on recent user data and then determines the threshold based on the communication budget of the cloud server and sends it to the edge. Edge inference refers to the MRS calculated when the click sequence on the edge is updated. The experimental results show that: 1) In terms of time consumption, both cloud preprocessing and edge inference are the fastest for random requests, followed by our IntellectReq. LOF and OC-SVM are the slowest. 2) In terms of space consumption, random, LOF, and OC-SVM can all be regarded as requiring no additional space consumption. In contrast, our method requires the additional deployment of 5.06k parameters on the edge. 3)  Random and our IntellectReq can be realized in terms of controllability. It means that edge-cloud communication can be realized under the condition of an arbitrary communication budget, while LOF and OC-SVM cannot. 4) In terms of high yield, LOF, OC-SVM, and our IntellectReq can all be achieved, but random requests cannot.
In general, our IntellectReq only requires minimal time consumption (does not affect real-time performance) and space consumption (easy to deploy for smart edges) and can take into account controllability and high profitability.

%% file: tex/5conclusion.tex
\section{Conclusion}
In our paper, we argue that under the EC-CDR framework, most communications requesting new parameters for the cloud-based recommendation system are unnecessary due to stable on-edge data distributions. We introduced IntellectReq, a low-resource solution for calculating request value and ensuring adaptive, high-revenue edge-cloud communication. IntellectReq employs a novel edge intelligence task to identify out-of-domain data and uses real-time user behavior mapping to a normal distribution, alongside multi-sampling outputs, to assess the edge model's adaptability to user actions. Our extensive tests across three public benchmarks confirm IntellectReq's efficiency and broad applicability, promoting a more effective edge-cloud collaborative recommendation approach.

%% file: tex/7acknowledge.tex
\subsection*{ACKNOWLEDGMENT}
This work was supported by National Key R\&D Program of China (No. 2022ZD0119100), Scientific Research Fund of Zhejiang Provincial Education Department (Y202353679), National Natural Science Foundation of China (No. 62376243, 62037001, U20A20387), the StarryNight Science Fund of Zhejiang University Shanghai Institute for Advanced Study (SN-ZJU-SIAS-0010), Project by Shanghai AI Laboratory (P22KS00111) and Program of Zhejiang Province Science and Technology (2022C01044)

%% file: tex/99appendix.tex
\appendix
\section{Appendix}
\label{sec:appendix}
This is the Appendix for ``Intelligent Model Update Strategy for Sequential Recommendation''.

\subsection{Supplementary Method}
\subsubsection{Notations and Definitions}
We summarize notations and definitions in the Table~\ref{tab:notation}.
\begin{table*}[!h]
\caption{Notations and Definitions}
\label{tab:notation}
\vspace{-0.4cm}
\resizebox{0.85\textwidth}{!}{
\begin{tabular}{c|c}
\toprule[2pt]
\textbf{Notation} & \textbf{Definition} \\
\midrule\midrule
$u$ & User \\
\rowcolor[HTML]{F8F8F8} $v$ & Item \\
$s$ & Behavior sequence \\
\rowcolor[HTML]{F8F8F8} $d$ & Edge \\
$\mathcal{D}=\{d^{(i)}\}_{i=1}^{\mathcal{N}_d}$ & Set of edges \\
\rowcolor[HTML]{F8F8F8} $\mathcal{S}_{H^{(i)}}$, $\mathcal{S}_{R^{(i)}}$, $\mathcal{S}_{MRD}$ & History samples, Real-time samples, MRD samples \\
$\mathcal{N}_{d}$, $\mathcal{N}_{H^{(i)}}$ and $\mathcal{N}_{R^{(i)}}$ & The number of edges, The number of history data, The number of real-time data \\
\rowcolor[HTML]{F8F8F8} $\Theta_g$, $\Theta_d$, $\Theta_{MRD}$ & Parameters of the global cloud model, Parameters of the local edge model \\
$\mathcal{M}_g(\cdot;\Theta_g)$, $\mathcal{M}_{d^{(i)}}(\cdot;\Theta_{d^{(i)}})$, $\mathcal{M}_{c^{(i)}_{t}}(\mathcal{S}_{_{\rm{MRD}}};\Theta_{_{\rm{MRD}}})$ & Global cloud model, Local edge recommendation model, Local edge control model \\
\rowcolor[HTML]{F8F8F8} $L_{rec}$, $L_{MRD}$ & Loss function of recommendation, Loss function of mis-recommendation \\
$\Omega$ & Feature extractor \\
\bottomrule[2pt]
\end{tabular}
}
\vspace{-0.2cm}
\end{table*}

\subsubsection{Optimization Target}

 To describe it in the simplest way, we assume that the set of the edges is $\mathcal{D}=\{d^{(i)}\}_{i=1}^{\mathcal{N}_d}$, the set updated using the baseline method is $\mathcal{D}'_u=\{d^{(i)}\}_{i=1}^{\mathcal{N}'_u}$, the set updated using our method is $\mathcal{D}_u=\{d^{(i)}\}_{i=1}^{\mathcal{N}_u}$. $\mathcal{N}_d$, $\mathcal{N}'_u$, and $\mathcal{N}_u$ are the amount of the $\mathcal{D}$, $\mathcal{D}'_u$ and $\mathcal{D}_u$, respectively. The communication upper bound is set to $N_{\rm{thres}}$. Suppose the ground-truth value $y$,  and the prediction of the baseline models $\hat{y}'$, and the prediction of our model $\hat{y}$ are row vectors.
Therefore, our optimization target is to obtain the highest performance of the model while limiting the upper bound of the communication frequency.
\begin{equation}
    \begin{aligned}
        \textbf{Maximize} \quad & \hat{y}y^T,\\
        \textbf{Subject to} \quad & 0 \leq \mathcal{N}_u \leq N_{\rm{thres}},\\
        &  \mathcal{N}_u \leq \mathcal{N}'_u,\\
        & \mathcal{D}_u\subset\mathcal{D}.
    \end{aligned}
\end{equation}
In this case, the improvement of our method is $\Delta=\hat{y}y^T-\hat{y}'y^T$.

Or it can also be regarded as reducing the communication frequency without degrading performance.
\begin{equation}
    \begin{aligned}
        \textbf{Minimize} \quad &\mathcal{N}_u \\
        \textbf{Subject to} \quad & 0 \leq \mathcal{N}_u \leq N_{\rm{thres}}, \\
        & \hat{y}y^T \geq \hat{y}'y^T, \\
        & \mathcal{D}_u\subset\mathcal{D}
\end{aligned}
\end{equation}
In this case, the improvement of our method is $\Delta=\mathcal{N}-\mathcal{N}_u$.

\subsection{Supplementary Experimental Results}
\label{sec:appendix_supplementary_exp}

\subsubsection{Datasets.}
\begin{sloppypar}
We evaluate IntellectReq and baselines on \texttt{Amazon CDs~(CDs)}~\footnote{{https://jmcauley.ucsd.edu/data/amazon/}\label{fn:amazon}}, \texttt{Amazon Electronic~(Electronic)}~\footref{fn:amazon}, \texttt{Douban Book~(Book)}~\footnote{{https://www.kaggle.com/datasets/fengzhujoey/douban-datasetratingreviewside-information}\label{fn:douban}}, three widely used public benchmarks in the recommendation tasks, Table~\ref{tab:datasets_statistics} shows the statistics. Following conventional practice, all user-item pairs in the dataset are treated as positive samples. To conduct sequential recommendation experiments, we arrange the items clicked by the user into a sequence in the order of timestamps. 
We also refer to ~\cite{ref:din,ref:sasrec,ref:gru4rec}, which is negatively sampled at $1:4$ and $1:99$ in the training set and testing set, respectively. Negative sampling considers all user-item pairs that do not exist in the dataset as negative samples. 
\end{sloppypar}
\begin{table}[!h]
\centering
\caption{Statistics of Datasets.}
\vspace{-0.4cm}
\label{tab:datasets_statistics}
\resizebox{0.475\textwidth}{!}{
\begin{tabular}{c|c|c|c}
\toprule[1pt] 
\textbf{}     & \textbf{Amazon CDs} & \textbf{Amazon Electronic} & \textbf{Douban Books} \\
\midrule \midrule 
\#User        & 1,578,597           & 4,201,696                  & 46,549                        \\
\rowcolor[HTML]{F8F8F8} 
\#Item        & 486,360             & 476,002                    & 212,996                      \\
\#Interaction & 3,749,004           & 7,824,482                  & 1,861,533              \\
\rowcolor[HTML]{F8F8F8} 
\#Density     & 0.0000049           & 0.0000039                  & 0.0002746              \\
\bottomrule[1pt]     
\end{tabular}
}
\vspace{-0.3cm}
\end{table}

\subsubsection{Evaluation Metrics}
In the experiments, we use the widely adopted AUC, Logloss, HitRate and NDCG as the metrics to evaluate model performance. 
 They are defined by the following equations.
\begin{equation}
 \begin{aligned}
        \mathrm{AUC}=\frac{\sum_{x_0\in \mathcal{D}_T} \sum_{x_1 \in \mathcal{D}_F}\mathds{1}[f(x_1)<f(x_0)]}{|\mathcal{D}_T||\mathcal{D}_F|}, 
\end{aligned}
\end{equation}
\begin{equation}
\begin{aligned}
        \mathrm{UAUC}=\frac{1}{|\mathcal{U}|}\sum_{u\in \mathcal{U}}\frac{\sum_{x_0\in \mathcal{D}^u_T} \sum_{x_1 \in \mathcal{D}^u_F}\mathds{1}[f(x_1)<f(x_0)]}{|\mathcal{D}^u_T||\mathcal{D}^u_F|}, 
\end{aligned}
\end{equation}
\begin{equation}
\begin{aligned}
        \text{NDCG}@K = \sum_{u\in \mathcal{U}}  \frac 1 {|\mathcal{U}|}  \frac{2^{\mathds{1}(R_{u,g_u}\leq K)}-1}{\log_2(\mathds{1}(R_{u,g_u}\leq K)+1)},
\end{aligned}
\end{equation}
\begin{equation}
\begin{aligned}
        \text{HitRate}@K = \frac{1}{|\mathcal{U}|}\sum_{u\in \mathcal{U}} \mathds{1}(R_{u,g_u}\leq K), 
\end{aligned}
\end{equation}
    In the equation above, $\mathds{1}(\cdot)$ is the indicator function. $f$ is the model to be evaluated. $R_{u,g_u}$ is the rank predicted by the model for the ground truth item $g_u$ and user $u$. $\mathcal{D}_T$, $\mathcal{D}_F$ is the positive and negative testing sample set, respectively, and $\mathcal{D}^u_T$, $\mathcal{D}^u_F$ is the positive and negative testing sample set for user $u$ respectively.

\subsubsection{Request Frequency and Threshold}
\begin{figure}[!h]
\vspace{-0.3cm}
    \centering
    \includegraphics[width=0.95\linewidth]{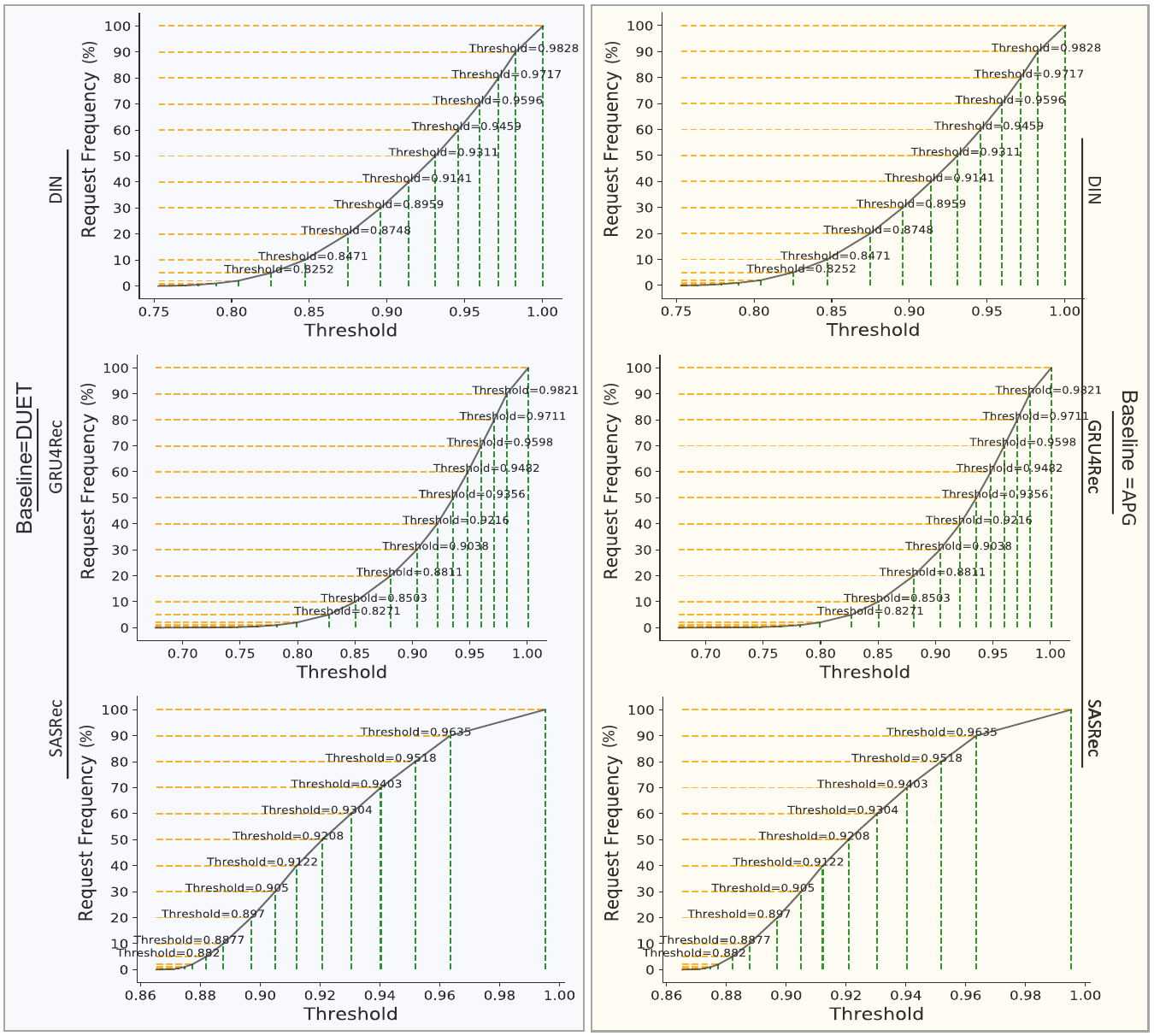}
    \vspace{-0.3cm}
    \caption{Request frequency \textit{w.r.t.} different threshold}
    \label{fig:threshold}
\vspace{-0.4cm}
\end{figure}

Figure~\ref{fig:threshold} shows that the relationship between request frequency and different threshold.

\subsection{Training Procedure and Inference Procedure}
\label{sec:training_inference_procedure}
\begin{figure*}[!h]
  \centering
\includegraphics[width=0.9\linewidth]{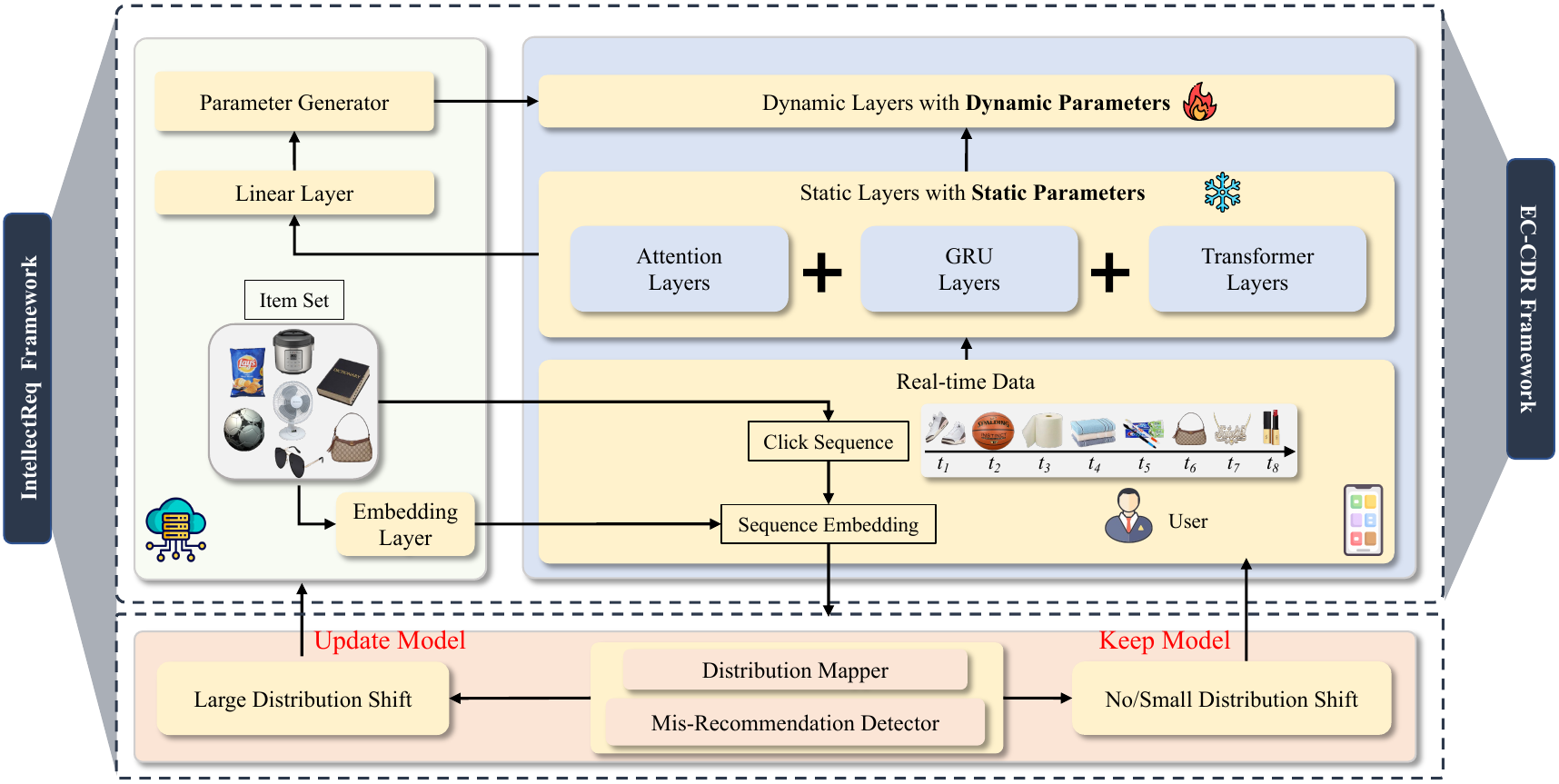}
\vspace{-0.3cm}
  \caption{The overall pipeline of our proposed IntellectReq.}
  \label{appendix:fig:overall_pipeline}
\end{figure*}

In this section, we describe the overall pipeline in detail in conjunction with Figure~\ref{appendix:fig:overall_pipeline}. 
\label{sec:train_inference_procedure}

1. \textbf{Training Procedure}

\textcircled{1} We first pre-trained a EC-CDR framework, and EC-CDR can use data to generate model parameters.

\textcircled{2} \textbf{MRD training procedure}. 1) \textbf{Construct the MRD dataset}. We assume that the time at this time is $T$, and then we use the model parameters generated by the data at moment $t=0$ under the EC-CDR framework, and the model is applied to the data at the current moment $t=T$. At this point, we can get a prediction result $\hat{y}$, compare $\hat{y}$ with $y$ to get whether the model do mis-recommendation. We then repeat the data used for parameter generation from $t=0$ to $t=T-1$, which constructs an MRD dataset. It contains three columns, namely: the data used for parameter generation ($x_1$), the current data ($x_2$), and whether it do mis-recommendation ($y_{\rm{MRD}}$). 2) \textbf{Train MRD}. MRD is a fully connected neural network that takes $x_1$ and $x_2$ as input and fits the mis-recommendation label $y_{\rm{MRD}}$. And then we get the MRD. MRD can be used to determine whether the model parameters generated using the data at a certain moment before are still valid for the current data. The prediction result output by MRD can be simply considered as Mis-Recommendation Score (MRS).

\textcircled{3} \textbf{DM training procedure}. We map the data into a Gaussian distribution through the Conditional-VAE method, and then sample the feature vector from the distribution to complete the next-item prediction task, that is, to predict the item that the user will click next. Then we can get DM. DM can calculate multiple next-items by sampling from the distribution multiple times, which can be used to calculate Uncertainty.

\textcircled{4} \textbf{Joint training procedure of MRD and DM}. We use a fully connected neural network, denoted as $f(\cdot)$, and use MRS and Uncertainty as input to fit $y_{\rm{MRD}}$ in the MRD dataset, which is the Mis-Recommendation Label.

2. \textbf{Inference Procedure}

The MRS is calculated using all recent user data on the cloud, and the threshold of the MRS is determined according to the load. Then send this threshold to each edge. The edge has updated the model at a certain moment $t=n, n<T$ before, and now whether it is necessary to continue to update the model at moment $ t=T $, that is, whether the model is invalid for the current data distribution? We only need to input the MRD and Uncertainty calculated by the data at the moment $t=n$ data and the data at the moment $t=T$ into $f(\cdot)$ for determine. In fact, what we output is a invalid degree, which is a continuous value between 0 and 1. Whether to update the edge model depends on the threshold calculated on the cloud based on the load.

\subsection{Hyperparameters and Training Schedules}
\label{sec:appendix_implementation_detail}

We summarize the hyperparameters and training schedules of \method{} on the three datasets in Table~\ref{tab:hyperparameters_and_training_schedule}.

\begin{table}[!h]
\vspace{-0.2cm}
    \caption{Hyperparameters and training schedules.}
    \centering
    \vspace{-0.3cm}
 \resizebox{0.35\textwidth}{!}{
    \begin{tabular}{c|c|c}
    \toprule[2pt]
    Dataset & Parameters & Setting \\ 
    \midrule
    \midrule
    \multirow{8}{*}{\makecell[c]{Amazon CDs\\Amazon Electronic\\Douban Book}} & GPU & Tesla A100 \\ \cline{2-3}
    \multirow{8}{*}{} & Optimizer & Adam\\ \cline{2-3}
    \multirow{8}{*}{} & \makecell[c]{Learning rate} & 0.001\\ \cline{2-3}
    \multirow{8}{*}{} & \makecell[c]{Batch size} & 1024 \\ \cline{2-3}
    \multirow{8}{*}{} & \makecell[c]{Sequence length} & 30 \\ \cline{2-3}
    \multirow{8}{*}{} & \makecell[c]{the Dimension of $z$} & 1×64 \\ \cline{2-3}
    \multirow{8}{*}{} & $N$ & 32 \\ \cline{2-3}
    \multirow{8}{*}{} & $n$ & 10 \\ 
     \bottomrule[2pt]
    \end{tabular}
   }
    \label{tab:hyperparameters_and_training_schedule}
    \vspace{-0.3cm}
\end{table}

\subsubsection{Impact on the Real World.}
A case based on a dynamic model from the previous moment is as follows. If it were based on a on-edge static model, the improvement would be much more significant.
We found some more intuitive data and examples to show the challenge and IntellectReq's impact on the real world: 
\begin{table}[!h]
\vspace{-0.2cm}
\caption{IntellectReq's Impact on Real World.}
\label{tab:real_world}
\vspace{-0.3cm}
\centering
\resizebox{0.35\textwidth}{!}{
\begin{tabular}{c|c|c|c|c}
\toprule[2pt]
 & \multicolumn{2}{c|}{Google} & \multicolumn{2}{c}{Alibaba} \\ \cline{2-5}
 & Bytes & FLOPs & Bytes & FLOPs \\ \midrule \midrule
EC-CDR & 4.69GB & 152.46G & 53.19GB & 1.68T \\ \cline{1-5}
IntellectReq & 3.79GB & 123.49G & 43.08GB & 1.36T \\ \cline{1-5}
$\Delta$ & \multicolumn{4}{c}{19.2\%} \\
\bottomrule[2pt]
\end{tabular}
}
\vspace{-0.2cm}
\end{table}
(1) We calculate the number of bytes and FLOPs required to update a parameter. Bytes: 48.5kB, FLOPs: 1.53M. That is, updating a model on the edge requires the transmission of 48.5kB data through edge-cloud communication, and consumes 1.53M computing power of the cloud model. (2) According to the report, Google processes 99,000 clicks per second, so it needs to transmit 48.5kB$*$99k=4.69GB per second, and consume 1.53M$*$99k=152.46G of computing power in the cloud server. Alibaba processes 1,150,000 clicks per second, so it needs to transmit 48.5kB$*$1150k=53.19GB per second, and consume 1.53M$*$1150k=1.68T of computing power in the cloud server. These are not the peak value yet. Obviously, such a huge loan and computing power consumption make it hard to update the model for edges every moment especially at peak times. (3) Sometimes, the distributed nature of clouds today may can afford the computational volume, since it can call enough servers to support edge-cloud collaboration. However, the huge resource  consumption is impractical in real-scenario. Besides, according to our empirical study, our IntellectReq can bring 21.4\% resource saving when the performance is the same using the APG framework. Under the DUET framework, IntellectReq can bring 16.6\% resource saving when the performance is the same. Summing up, IntellectReq can save 19\% resources on average, which is very helpful for cost control and can facilitate the EC-CDR development in practice. The following Table~\ref{tab:real_world} is the comparison between our method IntellectReq and EC-CDR in the amount of transmitted data and the computing power consumed on the cloud. (4) During the peak period, resources will be tight and cause freezes or even crashes. This is still in the case that EC-CDR has not been deployed yet, that is, the edge-cloud communication only performs the most basic user data transmission. Then, IntellectReq can achieve better performance than EC-CDR under any resource limit $\epsilon$, or to achieve the performance that EC-CDR requires $\epsilon+19\%$ of resources to achieve. 

%% file: main.bbl

\begin{thebibliography}{65}


\ifx \showCODEN    \undefined \def \showCODEN     #1{\unskip}     \fi
\ifx \showDOI      \undefined \def \showDOI       #1{#1}\fi
\ifx \showISBNx    \undefined \def \showISBNx     #1{\unskip}     \fi
\ifx \showISBNxiii \undefined \def \showISBNxiii  #1{\unskip}     \fi
\ifx \showISSN     \undefined \def \showISSN      #1{\unskip}     \fi
\ifx \showLCCN     \undefined \def \showLCCN      #1{\unskip}     \fi
\ifx \shownote     \undefined \def \shownote      #1{#1}          \fi
\ifx \showarticletitle \undefined \def \showarticletitle #1{#1}   \fi
\ifx \showURL      \undefined \def \showURL       {\relax}        \fi
\providecommand\bibfield[2]{#2}
\providecommand\bibinfo[2]{#2}
\providecommand\natexlab[1]{#1}
\providecommand\showeprint[2][]{arXiv:#2}

\bibitem[\protect\citeauthoryear{Breunig, Kriegel, Ng, and Sander}{Breunig et~al\mbox{.}}{2000}]%
        {ref:lof}
\bibfield{author}{\bibinfo{person}{Markus~M Breunig}, \bibinfo{person}{Hans-Peter Kriegel}, \bibinfo{person}{Raymond~T Ng}, {and} \bibinfo{person}{J{\"o}rg Sander}.} \bibinfo{year}{2000}\natexlab{}.
\newblock \showarticletitle{LOF: identifying density-based local outliers}. In \bibinfo{booktitle}{\emph{Proceedings of the 2000 ACM SIGMOD international conference on Management of data}}. \bibinfo{pages}{93--104}.
\newblock


\bibitem[\protect\citeauthoryear{Cai, Gan, Zhu, and Han}{Cai et~al\mbox{.}}{2020}]%
        {ref:finetuning}
\bibfield{author}{\bibinfo{person}{Han Cai}, \bibinfo{person}{Chuang Gan}, \bibinfo{person}{Ligeng Zhu}, {and} \bibinfo{person}{Song Han}.} \bibinfo{year}{2020}\natexlab{}.
\newblock \showarticletitle{Tinytl: Reduce activations, not trainable parameters for efficient on-device learning}.
\newblock  (\bibinfo{year}{2020}).
\newblock


\bibitem[\protect\citeauthoryear{Cao, Zheng, Hassanzadeh, Lamba, Liu, and Liu}{Cao et~al\mbox{.}}{2023}]%
        {cao2023_10.1145/3604237.3626868}
\bibfield{author}{\bibinfo{person}{Defu Cao}, \bibinfo{person}{Yixiang Zheng}, \bibinfo{person}{Parisa Hassanzadeh}, \bibinfo{person}{Simran Lamba}, \bibinfo{person}{Xiaomo Liu}, {and} \bibinfo{person}{Yan Liu}.} \bibinfo{year}{2023}\natexlab{}.
\newblock \showarticletitle{Large Scale Financial Time Series Forecasting with Multi-faceted Model}. In \bibinfo{booktitle}{\emph{Proceedings of the Fourth ACM International Conference on AI in Finance}} (<conf-loc>, <city>Brooklyn</city>, <state>NY</state>, <country>USA</country>, </conf-loc>) \emph{(\bibinfo{series}{ICAIF '23})}. \bibinfo{publisher}{Association for Computing Machinery}, \bibinfo{address}{New York, NY, USA}, \bibinfo{pages}{472–480}.
\newblock
\showISBNx{9798400702402}
\urldef\tempurl%
\url{https://doi.org/10.1145/3604237.3626868}
\showDOI{\tempurl}


\bibitem[\protect\citeauthoryear{Chang, Gao, Zheng, Hui, Niu, Song, Jin, and Li}{Chang et~al\mbox{.}}{2021}]%
        {ref:surge}
\bibfield{author}{\bibinfo{person}{Jianxin Chang}, \bibinfo{person}{Chen Gao}, \bibinfo{person}{Yu Zheng}, \bibinfo{person}{Yiqun Hui}, \bibinfo{person}{Yanan Niu}, \bibinfo{person}{Yang Song}, \bibinfo{person}{Depeng Jin}, {and} \bibinfo{person}{Yong Li}.} \bibinfo{year}{2021}\natexlab{}.
\newblock \showarticletitle{Sequential recommendation with graph neural networks}. In \bibinfo{booktitle}{\emph{Proceedings of the 44th International ACM SIGIR Conference on Research and Development in Information Retrieval}}. \bibinfo{pages}{378--387}.
\newblock


\bibitem[\protect\citeauthoryear{Chen and Wang}{Chen and Wang}{2021}]%
        {chen2021multi}
\bibfield{author}{\bibinfo{person}{Zhengyu Chen} {and} \bibinfo{person}{Donglin Wang}.} \bibinfo{year}{2021}\natexlab{}.
\newblock \showarticletitle{Multi-Initialization Meta-Learning with Domain Adaptation}. In \bibinfo{booktitle}{\emph{ICASSP 2021-2021 IEEE International Conference on Acoustics, Speech and Signal Processing (ICASSP)}}. IEEE, \bibinfo{pages}{1390--1394}.
\newblock


\bibitem[\protect\citeauthoryear{Chen, Xiao, and Kuang}{Chen et~al\mbox{.}}{2022}]%
        {chen2022ba}
\bibfield{author}{\bibinfo{person}{Zhengyu Chen}, \bibinfo{person}{Teng Xiao}, {and} \bibinfo{person}{Kun Kuang}.} \bibinfo{year}{2022}\natexlab{}.
\newblock \showarticletitle{BA-GNN: On Learning Bias-Aware Graph Neural Network}. In \bibinfo{booktitle}{\emph{2022 IEEE 38th International Conference on Data Engineering (ICDE)}}. IEEE, \bibinfo{pages}{3012--3024}.
\newblock


\bibitem[\protect\citeauthoryear{Chen, Xiao, Kuang, Lv, Zhang, Yang, Lu, Yang, and Wu}{Chen et~al\mbox{.}}{2023}]%
        {chen2023learning_arxiv}
\bibfield{author}{\bibinfo{person}{Zhengyu Chen}, \bibinfo{person}{Teng Xiao}, \bibinfo{person}{Kun Kuang}, \bibinfo{person}{Zheqi Lv}, \bibinfo{person}{Min Zhang}, \bibinfo{person}{Jinluan Yang}, \bibinfo{person}{Chengqiang Lu}, \bibinfo{person}{Hongxia Yang}, {and} \bibinfo{person}{Fei Wu}.} \bibinfo{year}{2023}\natexlab{}.
\newblock \showarticletitle{Learning to Reweight for Graph Neural Network}.
\newblock \bibinfo{journal}{\emph{arXiv preprint arXiv:2312.12475}} (\bibinfo{year}{2023}).
\newblock


\bibitem[\protect\citeauthoryear{Chen, Xiao, Kuang, Lv, Zhang, Yang, Lu, Yang, and Wu}{Chen et~al\mbox{.}}{2024}]%
        {chen2023learning}
\bibfield{author}{\bibinfo{person}{Zhengyu Chen}, \bibinfo{person}{Teng Xiao}, \bibinfo{person}{Kun Kuang}, \bibinfo{person}{Zheqi Lv}, \bibinfo{person}{Min Zhang}, \bibinfo{person}{Jinluan Yang}, \bibinfo{person}{Chengqiang Lu}, \bibinfo{person}{Hongxia Yang}, {and} \bibinfo{person}{Fei Wu}.} \bibinfo{year}{2024}\natexlab{}.
\newblock \showarticletitle{Learning to Reweight for Generalizable Graph Neural Network}.
\newblock \bibinfo{journal}{\emph{Proceedings of the AAAI conference on artificial intelligence}} (\bibinfo{year}{2024}).
\newblock


\bibitem[\protect\citeauthoryear{Chen, Xu, and Wang}{Chen et~al\mbox{.}}{2021}]%
        {chen2021deep}
\bibfield{author}{\bibinfo{person}{Zhengyu Chen}, \bibinfo{person}{Ziqing Xu}, {and} \bibinfo{person}{Donglin Wang}.} \bibinfo{year}{2021}\natexlab{}.
\newblock \showarticletitle{Deep transfer tensor decomposition with orthogonal constraint for recommender systems}. In \bibinfo{booktitle}{\emph{Proceedings of the AAAI Conference on Artificial Intelligence}}, Vol.~\bibinfo{volume}{35}. \bibinfo{pages}{4010--4018}.
\newblock


\bibitem[\protect\citeauthoryear{Ha, Dai, and Le}{Ha et~al\mbox{.}}{2017}]%
        {ref:hypernetwork_pioneering1}
\bibfield{author}{\bibinfo{person}{David Ha}, \bibinfo{person}{Andrew Dai}, {and} \bibinfo{person}{Quoc~V Le}.} \bibinfo{year}{2017}\natexlab{}.
\newblock \showarticletitle{Hypernetworks}.
\newblock  (\bibinfo{year}{2017}).
\newblock


\bibitem[\protect\citeauthoryear{Hidasi, Karatzoglou, Baltrunas, and Tikk}{Hidasi et~al\mbox{.}}{2016}]%
        {ref:gru4rec}
\bibfield{author}{\bibinfo{person}{Bal{\'a}zs Hidasi}, \bibinfo{person}{Alexandros Karatzoglou}, \bibinfo{person}{Linas Baltrunas}, {and} \bibinfo{person}{Domonkos Tikk}.} \bibinfo{year}{2016}\natexlab{}.
\newblock \showarticletitle{Session-based recommendations with recurrent neural networks}.
\newblock \bibinfo{journal}{\emph{International Conference on Learning Representations 2016}} (\bibinfo{year}{2016}).
\newblock


\bibitem[\protect\citeauthoryear{Huang, Huang, Yang, Ren, Liu, Li, Ye, Liu, Yin, and Zhao}{Huang et~al\mbox{.}}{2023}]%
        {huang2023make}
\bibfield{author}{\bibinfo{person}{Rongjie Huang}, \bibinfo{person}{Jiawei Huang}, \bibinfo{person}{Dongchao Yang}, \bibinfo{person}{Yi Ren}, \bibinfo{person}{Luping Liu}, \bibinfo{person}{Mingze Li}, \bibinfo{person}{Zhenhui Ye}, \bibinfo{person}{Jinglin Liu}, \bibinfo{person}{Xiang Yin}, {and} \bibinfo{person}{Zhou Zhao}.} \bibinfo{year}{2023}\natexlab{}.
\newblock \showarticletitle{Make-an-audio: Text-to-audio generation with prompt-enhanced diffusion models}.
\newblock \bibinfo{journal}{\emph{arXiv preprint arXiv:2301.12661}} (\bibinfo{year}{2023}).
\newblock


\bibitem[\protect\citeauthoryear{Huang, Lam, Wang, Su, Yu, Ren, and Zhao}{Huang et~al\mbox{.}}{2022a}]%
        {DBLP:conf/ijcai/HuangL0S00Z22}
\bibfield{author}{\bibinfo{person}{Rongjie Huang}, \bibinfo{person}{Max W.~Y. Lam}, \bibinfo{person}{Jun Wang}, \bibinfo{person}{Dan Su}, \bibinfo{person}{Dong Yu}, \bibinfo{person}{Yi Ren}, {and} \bibinfo{person}{Zhou Zhao}.} \bibinfo{year}{2022}\natexlab{a}.
\newblock \showarticletitle{FastDiff: {A} Fast Conditional Diffusion Model for High-Quality Speech Synthesis}. In \bibinfo{booktitle}{\emph{{IJCAI}}}. \bibinfo{publisher}{ijcai.org}, \bibinfo{pages}{4157--4163}.
\newblock


\bibitem[\protect\citeauthoryear{Huang, Ren, Liu, Cui, and Zhao}{Huang et~al\mbox{.}}{2022b}]%
        {huang2022generspeech}
\bibfield{author}{\bibinfo{person}{Rongjie Huang}, \bibinfo{person}{Yi Ren}, \bibinfo{person}{Jinglin Liu}, \bibinfo{person}{Chenye Cui}, {and} \bibinfo{person}{Zhou Zhao}.} \bibinfo{year}{2022}\natexlab{b}.
\newblock \showarticletitle{Generspeech: Towards style transfer for generalizable out-of-domain text-to-speech}.
\newblock \bibinfo{journal}{\emph{Advances in Neural Information Processing Systems}}  \bibinfo{volume}{35} (\bibinfo{year}{2022}), \bibinfo{pages}{10970--10983}.
\newblock


\bibitem[\protect\citeauthoryear{Ji, Liang, Liao, Fei, and Feng}{Ji et~al\mbox{.}}{2023a}]%
        {ji2023partial}
\bibfield{author}{\bibinfo{person}{Wei Ji}, \bibinfo{person}{Renjie Liang}, \bibinfo{person}{Lizi Liao}, \bibinfo{person}{Hao Fei}, {and} \bibinfo{person}{Fuli Feng}.} \bibinfo{year}{2023}\natexlab{a}.
\newblock \showarticletitle{Partial Annotation-based Video Moment Retrieval via Iterative Learning}. In \bibinfo{booktitle}{\emph{Proceedings of the 31th ACM international conference on Multimedia}}.
\newblock


\bibitem[\protect\citeauthoryear{Ji, Liu, Zhang, Wei, and Wang}{Ji et~al\mbox{.}}{2023b}]%
        {ji2023online}
\bibfield{author}{\bibinfo{person}{Wei Ji}, \bibinfo{person}{Xiangyan Liu}, \bibinfo{person}{An Zhang}, \bibinfo{person}{Yinwei Wei}, {and} \bibinfo{person}{Xiang Wang}.} \bibinfo{year}{2023}\natexlab{b}.
\newblock \showarticletitle{Online Distillation-enhanced Multi-modal Transformer for Sequential Recommendation}. In \bibinfo{booktitle}{\emph{Proceedings of the 31th ACM international conference on Multimedia}}.
\newblock


\bibitem[\protect\citeauthoryear{Kang and McAuley}{Kang and McAuley}{2018}]%
        {ref:sasrec}
\bibfield{author}{\bibinfo{person}{Wang-Cheng Kang} {and} \bibinfo{person}{Julian McAuley}.} \bibinfo{year}{2018}\natexlab{}.
\newblock \showarticletitle{Self-attentive sequential recommendation}. In \bibinfo{booktitle}{\emph{2018 IEEE International Conference on Data Mining (ICDM)}}. IEEE, \bibinfo{pages}{197--206}.
\newblock


\bibitem[\protect\citeauthoryear{Latifi, Mauro, and Jannach}{Latifi et~al\mbox{.}}{2021}]%
        {latifi2021session}
\bibfield{author}{\bibinfo{person}{Sara Latifi}, \bibinfo{person}{Noemi Mauro}, {and} \bibinfo{person}{Dietmar Jannach}.} \bibinfo{year}{2021}\natexlab{}.
\newblock \showarticletitle{Session-aware recommendation: A surprising quest for the state-of-the-art}.
\newblock \bibinfo{journal}{\emph{Information Sciences}}  \bibinfo{volume}{573} (\bibinfo{year}{2021}), \bibinfo{pages}{291--315}.
\newblock


\bibitem[\protect\citeauthoryear{Li, Xiao, Zheng, Wu, and Cui}{Li et~al\mbox{.}}{2023e}]%
        {li2023propensity}
\bibfield{author}{\bibinfo{person}{Haoxuan Li}, \bibinfo{person}{Yanghao Xiao}, \bibinfo{person}{Chunyuan Zheng}, \bibinfo{person}{Peng Wu}, {and} \bibinfo{person}{Peng Cui}.} \bibinfo{year}{2023}\natexlab{e}.
\newblock \showarticletitle{Propensity matters: Measuring and enhancing balancing for recommendation}. In \bibinfo{booktitle}{\emph{International Conference on Machine Learning}}. PMLR, \bibinfo{pages}{20182--20194}.
\newblock


\bibitem[\protect\citeauthoryear{Li, Xiao, Zheng, Wu, Geng, Chen, and Cui}{Li et~al\mbox{.}}{2024}]%
        {li2024kernel}
\bibfield{author}{\bibinfo{person}{Haoxuan Li}, \bibinfo{person}{Yanghao Xiao}, \bibinfo{person}{Chunyuan Zheng}, \bibinfo{person}{Peng Wu}, \bibinfo{person}{Zhi Geng}, \bibinfo{person}{Xu Chen}, {and} \bibinfo{person}{Peng Cui}.} \bibinfo{year}{2024}\natexlab{}.
\newblock \showarticletitle{Debiased Collaborative Filtering with Kernel-based Causal Balancing}. In \bibinfo{booktitle}{\emph{International Conference on Learning Representations}}.
\newblock


\bibitem[\protect\citeauthoryear{Li, He, Wei, Qian, Zhu, Xie, Zhuang, Tian, and Tang}{Li et~al\mbox{.}}{2022a}]%
        {li2022fine}
\bibfield{author}{\bibinfo{person}{Juncheng Li}, \bibinfo{person}{Xin He}, \bibinfo{person}{Longhui Wei}, \bibinfo{person}{Long Qian}, \bibinfo{person}{Linchao Zhu}, \bibinfo{person}{Lingxi Xie}, \bibinfo{person}{Yueting Zhuang}, \bibinfo{person}{Qi Tian}, {and} \bibinfo{person}{Siliang Tang}.} \bibinfo{year}{2022}\natexlab{a}.
\newblock \showarticletitle{Fine-grained semantically aligned vision-language pre-training}.
\newblock \bibinfo{journal}{\emph{Advances in neural information processing systems}}  \bibinfo{volume}{35} (\bibinfo{year}{2022}), \bibinfo{pages}{7290--7303}.
\newblock


\bibitem[\protect\citeauthoryear{Li, Pan, Ge, Gao, Zhang, Ji, Zhang, Chua, Tang, and Zhuang}{Li et~al\mbox{.}}{2023a}]%
        {li2023finetuning}
\bibfield{author}{\bibinfo{person}{Juncheng Li}, \bibinfo{person}{Kaihang Pan}, \bibinfo{person}{Zhiqi Ge}, \bibinfo{person}{Minghe Gao}, \bibinfo{person}{Hanwang Zhang}, \bibinfo{person}{Wei Ji}, \bibinfo{person}{Wenqiao Zhang}, \bibinfo{person}{Tat-Seng Chua}, \bibinfo{person}{Siliang Tang}, {and} \bibinfo{person}{Yueting Zhuang}.} \bibinfo{year}{2023}\natexlab{a}.
\newblock \showarticletitle{Fine-tuning Multimodal LLMs to Follow Zero-shot Demonstrative Instructions}.
\newblock \bibinfo{journal}{\emph{arXiv preprint arXiv:2308.04152}} (\bibinfo{year}{2023}).
\newblock


\bibitem[\protect\citeauthoryear{Li, Wang, Qin, Ji, and Liang}{Li et~al\mbox{.}}{2023b}]%
        {lili_10.1145/3581783.3611847}
\bibfield{author}{\bibinfo{person}{Li Li}, \bibinfo{person}{Chenwei Wang}, \bibinfo{person}{You Qin}, \bibinfo{person}{Wei Ji}, {and} \bibinfo{person}{Renjie Liang}.} \bibinfo{year}{2023}\natexlab{b}.
\newblock \showarticletitle{Biased-Predicate Annotation Identification via Unbiased Visual Predicate Representation}. In \bibinfo{booktitle}{\emph{Proceedings of the 31st ACM International Conference on Multimedia}} (<conf-loc>, <city>Ottawa ON</city>, <country>Canada</country>, </conf-loc>) \emph{(\bibinfo{series}{MM '23})}. \bibinfo{publisher}{Association for Computing Machinery}, \bibinfo{address}{New York, NY, USA}, \bibinfo{pages}{4410–4420}.
\newblock
\showISBNx{9798400701085}
\urldef\tempurl%
\url{https://doi.org/10.1145/3581783.3611847}
\showDOI{\tempurl}


\bibitem[\protect\citeauthoryear{Li, Wang, Zhang, Miao, Zhao, Zhang, Ji, and Wu}{Li et~al\mbox{.}}{2023d}]%
        {li2023winner}
\bibfield{author}{\bibinfo{person}{Mengze Li}, \bibinfo{person}{Han Wang}, \bibinfo{person}{Wenqiao Zhang}, \bibinfo{person}{Jiaxu Miao}, \bibinfo{person}{Zhou Zhao}, \bibinfo{person}{Shengyu Zhang}, \bibinfo{person}{Wei Ji}, {and} \bibinfo{person}{Fei Wu}.} \bibinfo{year}{2023}\natexlab{d}.
\newblock \showarticletitle{Winner: Weakly-supervised hierarchical decomposition and alignment for spatio-temporal video grounding}. In \bibinfo{booktitle}{\emph{Proceedings of the IEEE/CVF Conference on Computer Vision and Pattern Recognition}}. \bibinfo{pages}{23090--23099}.
\newblock


\bibitem[\protect\citeauthoryear{Li, Wang, Xu, Han, Zhang, Zhao, Miao, Zhang, Pu, and Wu}{Li et~al\mbox{.}}{2023c}]%
        {li2023multi}
\bibfield{author}{\bibinfo{person}{Mengze Li}, \bibinfo{person}{Tianbao Wang}, \bibinfo{person}{Jiahe Xu}, \bibinfo{person}{Kairong Han}, \bibinfo{person}{Shengyu Zhang}, \bibinfo{person}{Zhou Zhao}, \bibinfo{person}{Jiaxu Miao}, \bibinfo{person}{Wenqiao Zhang}, \bibinfo{person}{Shiliang Pu}, {and} \bibinfo{person}{Fei Wu}.} \bibinfo{year}{2023}\natexlab{c}.
\newblock \showarticletitle{Multi-modal Action Chain Abductive Reasoning}. In \bibinfo{booktitle}{\emph{Proceedings of the 61st Annual Meeting of the Association for Computational Linguistics (Volume 1: Long Papers)}}. \bibinfo{pages}{4617--4628}.
\newblock


\bibitem[\protect\citeauthoryear{Li, Wang, Zhang, Zhang, Zhao, Miao, Zhang, Tan, Wang, Wang, et~al\mbox{.}}{Li et~al\mbox{.}}{2022b}]%
        {li2022end}
\bibfield{author}{\bibinfo{person}{Mengze Li}, \bibinfo{person}{Tianbao Wang}, \bibinfo{person}{Haoyu Zhang}, \bibinfo{person}{Shengyu Zhang}, \bibinfo{person}{Zhou Zhao}, \bibinfo{person}{Jiaxu Miao}, \bibinfo{person}{Wenqiao Zhang}, \bibinfo{person}{Wenming Tan}, \bibinfo{person}{Jin Wang}, \bibinfo{person}{Peng Wang}, {et~al\mbox{.}}} \bibinfo{year}{2022}\natexlab{b}.
\newblock \showarticletitle{End-to-End Modeling via Information Tree for One-Shot Natural Language Spatial Video Grounding}. In \bibinfo{booktitle}{\emph{Proceedings of the 60th Annual Meeting of the Association for Computational Linguistics (Volume 1: Long Papers)}}. \bibinfo{pages}{8707--8717}.
\newblock


\bibitem[\protect\citeauthoryear{Lin, Xu, Wang, Zhang, and Feng}{Lin et~al\mbox{.}}{2023}]%
        {lin2023mitigating}
\bibfield{author}{\bibinfo{person}{Xin-Yu Lin}, \bibinfo{person}{Yi-Yan Xu}, \bibinfo{person}{Wen-Jie Wang}, \bibinfo{person}{Yang Zhang}, {and} \bibinfo{person}{Fu-Li Feng}.} \bibinfo{year}{2023}\natexlab{}.
\newblock \showarticletitle{Mitigating Spurious Correlations for Self-supervised Recommendation}.
\newblock \bibinfo{journal}{\emph{Machine Intelligence Research}} \bibinfo{volume}{20}, \bibinfo{number}{2} (\bibinfo{year}{2023}), \bibinfo{pages}{263--275}.
\newblock


\bibitem[\protect\citeauthoryear{Lv, Wang, Zhang, Kuang, Yang, and Wu}{Lv et~al\mbox{.}}{2022}]%
        {lv2022personalizing}
\bibfield{author}{\bibinfo{person}{Zheqi Lv}, \bibinfo{person}{Feng Wang}, \bibinfo{person}{Shengyu Zhang}, \bibinfo{person}{Kun Kuang}, \bibinfo{person}{Hongxia Yang}, {and} \bibinfo{person}{Fei Wu}.} \bibinfo{year}{2022}\natexlab{}.
\newblock \showarticletitle{Personalizing Intervened Network for Long-tailed Sequential User Behavior Modeling}.
\newblock \bibinfo{journal}{\emph{arXiv preprint arXiv:2208.09130}} (\bibinfo{year}{2022}).
\newblock


\bibitem[\protect\citeauthoryear{Lv, Wang, Zhang, Zhang, Kuang, and Wu}{Lv et~al\mbox{.}}{2023a}]%
        {lv2023parameters}
\bibfield{author}{\bibinfo{person}{Zheqi Lv}, \bibinfo{person}{Feng Wang}, \bibinfo{person}{Shengyu Zhang}, \bibinfo{person}{Wenqiao Zhang}, \bibinfo{person}{Kun Kuang}, {and} \bibinfo{person}{Fei Wu}.} \bibinfo{year}{2023}\natexlab{a}.
\newblock \showarticletitle{Parameters Efficient Fine-Tuning for Long-Tailed Sequential Recommendation}. In \bibinfo{booktitle}{\emph{CAAI International Conference on Artificial Intelligence}}. Springer, \bibinfo{pages}{442--459}.
\newblock


\bibitem[\protect\citeauthoryear{Lv, Zhang, Zhang, Kuang, Wang, Wang, Chen, Shen, Yang, Ooi, and Wu}{Lv et~al\mbox{.}}{2023b}]%
        {ref:duet}
\bibfield{author}{\bibinfo{person}{Zheqi Lv}, \bibinfo{person}{Wenqiao Zhang}, \bibinfo{person}{Shengyu Zhang}, \bibinfo{person}{Kun Kuang}, \bibinfo{person}{Feng Wang}, \bibinfo{person}{Yongwei Wang}, \bibinfo{person}{Zhengyu Chen}, \bibinfo{person}{Tao Shen}, \bibinfo{person}{Hongxia Yang}, \bibinfo{person}{Beng~Chin Ooi}, {and} \bibinfo{person}{Fei Wu}.} \bibinfo{year}{2023}\natexlab{b}.
\newblock \showarticletitle{DUET: A Tuning-Free Device-Cloud Collaborative Parameters Generation Framework for Efficient Device Model Generalization}. In \bibinfo{booktitle}{\emph{Proceedings of the ACM Web Conference 2023}}.
\newblock


\bibitem[\protect\citeauthoryear{Marfoq, Neglia, Bellet, Kameni, and Vidal}{Marfoq et~al\mbox{.}}{2021}]%
        {ref:federated_multi_task2}
\bibfield{author}{\bibinfo{person}{Othmane Marfoq}, \bibinfo{person}{Giovanni Neglia}, \bibinfo{person}{Aur{\'e}lien Bellet}, \bibinfo{person}{Laetitia Kameni}, {and} \bibinfo{person}{Richard Vidal}.} \bibinfo{year}{2021}\natexlab{}.
\newblock \showarticletitle{Federated multi-task learning under a mixture of distributions}.
\newblock \bibinfo{journal}{\emph{Advances in Neural Information Processing Systems}}  \bibinfo{volume}{34} (\bibinfo{year}{2021}), \bibinfo{pages}{15434--15447}.
\newblock


\bibitem[\protect\citeauthoryear{McMahan, Moore, Ramage, Hampson, and y~Arcas}{McMahan et~al\mbox{.}}{2017}]%
        {ref:federated_fedavg}
\bibfield{author}{\bibinfo{person}{Brendan McMahan}, \bibinfo{person}{Eider Moore}, \bibinfo{person}{Daniel Ramage}, \bibinfo{person}{Seth Hampson}, {and} \bibinfo{person}{Blaise~Aguera y Arcas}.} \bibinfo{year}{2017}\natexlab{}.
\newblock \showarticletitle{Communication-efficient learning of deep networks from decentralized data}. In \bibinfo{booktitle}{\emph{Artificial intelligence and statistics}}. PMLR, \bibinfo{pages}{1273--1282}.
\newblock


\bibitem[\protect\citeauthoryear{Mills, Hu, and Min}{Mills et~al\mbox{.}}{2021}]%
        {ref:federated_multi_task}
\bibfield{author}{\bibinfo{person}{Jed Mills}, \bibinfo{person}{Jia Hu}, {and} \bibinfo{person}{Geyong Min}.} \bibinfo{year}{2021}\natexlab{}.
\newblock \showarticletitle{Multi-task federated learning for personalised deep neural networks in edge computing}.
\newblock \bibinfo{journal}{\emph{IEEE Transactions on Parallel and Distributed Systems}} \bibinfo{volume}{33}, \bibinfo{number}{3} (\bibinfo{year}{2021}), \bibinfo{pages}{630--641}.
\newblock


\bibitem[\protect\citeauthoryear{Qian, Xu, Lv, Zhang, Jiang, Liu, Zeng, Chua, and Wu}{Qian et~al\mbox{.}}{2022}]%
        {zhangsyDBLP:conf/kdd/QianXLZJLZC022}
\bibfield{author}{\bibinfo{person}{Xufeng Qian}, \bibinfo{person}{Yue Xu}, \bibinfo{person}{Fuyu Lv}, \bibinfo{person}{Shengyu Zhang}, \bibinfo{person}{Ziwen Jiang}, \bibinfo{person}{Qingwen Liu}, \bibinfo{person}{Xiaoyi Zeng}, \bibinfo{person}{Tat{-}Seng Chua}, {and} \bibinfo{person}{Fei Wu}.} \bibinfo{year}{2022}\natexlab{}.
\newblock \showarticletitle{Intelligent Request Strategy Design in Recommender System}. In \bibinfo{booktitle}{\emph{{KDD} '22: The 28th {ACM} {SIGKDD} Conference on Knowledge Discovery and Data Mining}}. \bibinfo{publisher}{{ACM}}, \bibinfo{pages}{3772--3782}.
\newblock


\bibitem[\protect\citeauthoryear{Qin, Lv, Wang, Hu, and Wu}{Qin et~al\mbox{.}}{2020}]%
        {qin2020health}
\bibfield{author}{\bibinfo{person}{Fang-Yu Qin}, \bibinfo{person}{Zhe-Qi Lv}, \bibinfo{person}{Dan-Ni Wang}, \bibinfo{person}{Bo Hu}, {and} \bibinfo{person}{Chao Wu}.} \bibinfo{year}{2020}\natexlab{}.
\newblock \showarticletitle{Health status prediction for the elderly based on machine learning}.
\newblock \bibinfo{journal}{\emph{Archives of gerontology and geriatrics}}  \bibinfo{volume}{90} (\bibinfo{year}{2020}), \bibinfo{pages}{104121}.
\newblock


\bibitem[\protect\citeauthoryear{Rendle, Freudenthaler, and Schmidt-Thieme}{Rendle et~al\mbox{.}}{2010}]%
        {ref:fpmc}
\bibfield{author}{\bibinfo{person}{Steffen Rendle}, \bibinfo{person}{Christoph Freudenthaler}, {and} \bibinfo{person}{Lars Schmidt-Thieme}.} \bibinfo{year}{2010}\natexlab{}.
\newblock \showarticletitle{Factorizing personalized Markov chains for next-basket recommendation}.
\newblock \bibinfo{journal}{\emph{the web conference}} (\bibinfo{year}{2010}).
\newblock


\bibitem[\protect\citeauthoryear{Sanh, Debut, Chaumond, and Wolf}{Sanh et~al\mbox{.}}{2019}]%
        {ref:disitll}
\bibfield{author}{\bibinfo{person}{Victor Sanh}, \bibinfo{person}{Lysandre Debut}, \bibinfo{person}{Julien Chaumond}, {and} \bibinfo{person}{Thomas Wolf}.} \bibinfo{year}{2019}\natexlab{}.
\newblock \showarticletitle{DistilBERT, a distilled version of BERT: smaller, faster, cheaper and lighter}.
\newblock \bibinfo{journal}{\emph{arXiv preprint arXiv:1910.01108}} (\bibinfo{year}{2019}).
\newblock


\bibitem[\protect\citeauthoryear{Su, Chen, Lin, Li, Liu, and Zheng}{Su et~al\mbox{.}}{2023a}]%
        {su2023personalized}
\bibfield{author}{\bibinfo{person}{Jiajie Su}, \bibinfo{person}{Chaochao Chen}, \bibinfo{person}{Zibin Lin}, \bibinfo{person}{Xi Li}, \bibinfo{person}{Weiming Liu}, {and} \bibinfo{person}{Xiaolin Zheng}.} \bibinfo{year}{2023}\natexlab{a}.
\newblock \showarticletitle{Personalized Behavior-Aware Transformer for Multi-Behavior Sequential Recommendation}. In \bibinfo{booktitle}{\emph{Proceedings of the 31st ACM International Conference on Multimedia}}. \bibinfo{pages}{6321--6331}.
\newblock


\bibitem[\protect\citeauthoryear{Su, Chen, Liu, Wu, Zheng, and Lyu}{Su et~al\mbox{.}}{2023b}]%
        {su2023enhancing}
\bibfield{author}{\bibinfo{person}{Jiajie Su}, \bibinfo{person}{Chaochao Chen}, \bibinfo{person}{Weiming Liu}, \bibinfo{person}{Fei Wu}, \bibinfo{person}{Xiaolin Zheng}, {and} \bibinfo{person}{Haoming Lyu}.} \bibinfo{year}{2023}\natexlab{b}.
\newblock \showarticletitle{Enhancing Hierarchy-Aware Graph Networks with Deep Dual Clustering for Session-based Recommendation}. In \bibinfo{booktitle}{\emph{Proceedings of the ACM Web Conference 2023}}. \bibinfo{pages}{165--176}.
\newblock


\bibitem[\protect\citeauthoryear{Sun, Liu, Wu, Pei, Lin, Ou, and Jiang}{Sun et~al\mbox{.}}{2019}]%
        {ref:bert4rec}
\bibfield{author}{\bibinfo{person}{Fei Sun}, \bibinfo{person}{Jun Liu}, \bibinfo{person}{Jian Wu}, \bibinfo{person}{Changhua Pei}, \bibinfo{person}{Xiao Lin}, \bibinfo{person}{Wenwu Ou}, {and} \bibinfo{person}{Peng Jiang}.} \bibinfo{year}{2019}\natexlab{}.
\newblock \showarticletitle{BERT4Rec: Sequential recommendation with bidirectional encoder representations from transformer}. In \bibinfo{booktitle}{\emph{Proceedings of the 28th ACM international conference on information and knowledge management}}. \bibinfo{pages}{1441--1450}.
\newblock


\bibitem[\protect\citeauthoryear{Tang, Lv, Zhang, Wu, and Kuang}{Tang et~al\mbox{.}}{2024a}]%
        {tang2024modelgpt}
\bibfield{author}{\bibinfo{person}{Zihao Tang}, \bibinfo{person}{Zheqi Lv}, \bibinfo{person}{Shengyu Zhang}, \bibinfo{person}{Fei Wu}, {and} \bibinfo{person}{Kun Kuang}.} \bibinfo{year}{2024}\natexlab{a}.
\newblock \showarticletitle{ModelGPT: Unleashing LLM's Capabilities for Tailored Model Generation}.
\newblock \bibinfo{journal}{\emph{arXiv preprint arXiv:2402.12408}} (\bibinfo{year}{2024}).
\newblock


\bibitem[\protect\citeauthoryear{Tang, Lv, Zhang, Zhou, Duan, Kuang, and Wu}{Tang et~al\mbox{.}}{2024b}]%
        {tang2024oodkd}
\bibfield{author}{\bibinfo{person}{Zihao Tang}, \bibinfo{person}{Zheqi Lv}, \bibinfo{person}{Shengyu Zhang}, \bibinfo{person}{Yifan Zhou}, \bibinfo{person}{Xinyu Duan}, \bibinfo{person}{Kun Kuang}, {and} \bibinfo{person}{Fei Wu}.} \bibinfo{year}{2024}\natexlab{b}.
\newblock \showarticletitle{AuG-KD: Anchor-Based Mixup Generation for Out-of-Domain Knowledge Distillation}. In \bibinfo{booktitle}{\emph{12th International Conference on Learning Representations, {ICLR} 2024, Vienna Austria, May 7-11, 2024}}. \bibinfo{publisher}{OpenReview.net}.
\newblock
\urldef\tempurl%
\url{https://openreview.net/forum?id=fcqWJ8JgMR}
\showURL{%
\tempurl}


\bibitem[\protect\citeauthoryear{Tax}{Tax}{2002}]%
        {ref:ocsvm}
\bibfield{author}{\bibinfo{person}{David Martinus~Johannes Tax}.} \bibinfo{year}{2002}\natexlab{}.
\newblock \showarticletitle{One-class classification: Concept learning in the absence of counter-examples.}
\newblock  (\bibinfo{year}{2002}).
\newblock


\bibitem[\protect\citeauthoryear{Tong, Yuan, Zhang, Zhu, Zhang, Wu, and Kuang}{Tong et~al\mbox{.}}{2023}]%
        {DBLP:conf/kdd/TongYZZZWK23}
\bibfield{author}{\bibinfo{person}{Yunze Tong}, \bibinfo{person}{Junkun Yuan}, \bibinfo{person}{Min Zhang}, \bibinfo{person}{Didi Zhu}, \bibinfo{person}{Keli Zhang}, \bibinfo{person}{Fei Wu}, {and} \bibinfo{person}{Kun Kuang}.} \bibinfo{year}{2023}\natexlab{}.
\newblock \showarticletitle{Quantitatively Measuring and Contrastively Exploring Heterogeneity for Domain Generalization}. In \bibinfo{booktitle}{\emph{{KDD}}}. \bibinfo{publisher}{{ACM}}, \bibinfo{pages}{2189--2200}.
\newblock


\bibitem[\protect\citeauthoryear{Wang, Cui, Wang, Pei, Zhu, and Yang}{Wang et~al\mbox{.}}{2017}]%
        {wang2017community}
\bibfield{author}{\bibinfo{person}{Xiao Wang}, \bibinfo{person}{Peng Cui}, \bibinfo{person}{Jing Wang}, \bibinfo{person}{Jian Pei}, \bibinfo{person}{Wenwu Zhu}, {and} \bibinfo{person}{Shiqiang Yang}.} \bibinfo{year}{2017}\natexlab{}.
\newblock \showarticletitle{Community preserving network embedding}. In \bibinfo{booktitle}{\emph{Proceedings of the AAAI conference on artificial intelligence}}, Vol.~\bibinfo{volume}{31}.
\newblock


\bibitem[\protect\citeauthoryear{Wu, Tang, Zhu, Wang, Xie, and Tan}{Wu et~al\mbox{.}}{2019}]%
        {ref:srgnn}
\bibfield{author}{\bibinfo{person}{Shu Wu}, \bibinfo{person}{Yuyuan Tang}, \bibinfo{person}{Yanqiao Zhu}, \bibinfo{person}{Liang Wang}, \bibinfo{person}{Xing Xie}, {and} \bibinfo{person}{Tieniu Tan}.} \bibinfo{year}{2019}\natexlab{}.
\newblock \showarticletitle{Session-based recommendation with graph neural networks}. In \bibinfo{booktitle}{\emph{Proceedings of the AAAI conference on artificial intelligence}}, Vol.~\bibinfo{volume}{33}. \bibinfo{pages}{346--353}.
\newblock


\bibitem[\protect\citeauthoryear{Wu, Lu, Zhang, Jatowt, Feng, Sun, Wu, and Kuang}{Wu et~al\mbox{.}}{2023a}]%
        {wu2023focus}
\bibfield{author}{\bibinfo{person}{Yiquan Wu}, \bibinfo{person}{Weiming Lu}, \bibinfo{person}{Yating Zhang}, \bibinfo{person}{Adam Jatowt}, \bibinfo{person}{Jun Feng}, \bibinfo{person}{Changlong Sun}, \bibinfo{person}{Fei Wu}, {and} \bibinfo{person}{Kun Kuang}.} \bibinfo{year}{2023}\natexlab{a}.
\newblock \showarticletitle{Focus-aware response generation in inquiry conversation}. In \bibinfo{booktitle}{\emph{Findings of the Association for Computational Linguistics: ACL 2023}}. \bibinfo{pages}{12585--12599}.
\newblock


\bibitem[\protect\citeauthoryear{Wu, Zhou, Liu, Lu, Liu, Zhang, Sun, Wu, and Kuang}{Wu et~al\mbox{.}}{2023b}]%
        {wu2023precedent}
\bibfield{author}{\bibinfo{person}{Yiquan Wu}, \bibinfo{person}{Siying Zhou}, \bibinfo{person}{Yifei Liu}, \bibinfo{person}{Weiming Lu}, \bibinfo{person}{Xiaozhong Liu}, \bibinfo{person}{Yating Zhang}, \bibinfo{person}{Changlong Sun}, \bibinfo{person}{Fei Wu}, {and} \bibinfo{person}{Kun Kuang}.} \bibinfo{year}{2023}\natexlab{b}.
\newblock \showarticletitle{Precedent-Enhanced Legal Judgment Prediction with LLM and Domain-Model Collaboration}.
\newblock \bibinfo{journal}{\emph{arXiv preprint arXiv:2310.09241}} (\bibinfo{year}{2023}).
\newblock


\bibitem[\protect\citeauthoryear{Xinyu~Lin and Chua}{Xinyu~Lin and Chua}{2024}]%
        {lin2023temporally}
\bibfield{author}{\bibinfo{person}{Jujia Zhao Yongqi Li Fuli~Feng Xinyu~Lin, Wenjie~Wang} {and} \bibinfo{person}{Tat-Seng Chua}.} \bibinfo{year}{2024}\natexlab{}.
\newblock \showarticletitle{Temporally and Distributionally Robust Optimization for Cold-start Recommendation}. In \bibinfo{booktitle}{\emph{AAAI}}.
\newblock


\bibitem[\protect\citeauthoryear{Yan, Wang, Zhang, Li, Xu, and Zheng}{Yan et~al\mbox{.}}{2022b}]%
        {ref:apg_rs1}
\bibfield{author}{\bibinfo{person}{Bencheng Yan}, \bibinfo{person}{Pengjie Wang}, \bibinfo{person}{Kai Zhang}, \bibinfo{person}{Feng Li}, \bibinfo{person}{Jian Xu}, {and} \bibinfo{person}{Bo Zheng}.} \bibinfo{year}{2022}\natexlab{b}.
\newblock \showarticletitle{APG: Adaptive Parameter Generation Network for Click-Through Rate Prediction}. In \bibinfo{booktitle}{\emph{Advances in Neural Information Processing Systems}}.
\newblock


\bibitem[\protect\citeauthoryear{Yan, Niu, Gu, Wu, Tang, Hua, Lyu, and Chen}{Yan et~al\mbox{.}}{2022a}]%
        {ref:edge_cloud2}
\bibfield{author}{\bibinfo{person}{Yikai Yan}, \bibinfo{person}{Chaoyue Niu}, \bibinfo{person}{Renjie Gu}, \bibinfo{person}{Fan Wu}, \bibinfo{person}{Shaojie Tang}, \bibinfo{person}{Lifeng Hua}, \bibinfo{person}{Chengfei Lyu}, {and} \bibinfo{person}{Guihai Chen}.} \bibinfo{year}{2022}\natexlab{a}.
\newblock \showarticletitle{On-Device Learning for Model Personalization with Large-Scale Cloud-Coordinated Domain Adaption}. In \bibinfo{booktitle}{\emph{{KDD} '22: The 28th {ACM} {SIGKDD} Conference on Knowledge Discovery and Data Mining, Washington, DC, USA, August 14 - 18, 2022}}. \bibinfo{pages}{2180--2190}.
\newblock


\bibitem[\protect\citeauthoryear{Yao, Wang, Ding, Chen, Han, Zhou, and Yang}{Yao et~al\mbox{.}}{2022a}]%
        {ref:edge_cloud}
\bibfield{author}{\bibinfo{person}{Jiangchao Yao}, \bibinfo{person}{Feng Wang}, \bibinfo{person}{Xichen Ding}, \bibinfo{person}{Shaohu Chen}, \bibinfo{person}{Bo Han}, \bibinfo{person}{Jingren Zhou}, {and} \bibinfo{person}{Hongxia Yang}.} \bibinfo{year}{2022}\natexlab{a}.
\newblock \showarticletitle{Device-cloud Collaborative Recommendation via Meta Controller}. In \bibinfo{booktitle}{\emph{{KDD} '22: The 28th {ACM} {SIGKDD} Conference on Knowledge Discovery and Data Mining, Washington, DC, USA, August 14 - 18, 2022}}. \bibinfo{pages}{4353--4362}.
\newblock


\bibitem[\protect\citeauthoryear{Yao, Zhang, Yao, Wang, Ma, Zhang, Chu, Ji, Jia, Shen, et~al\mbox{.}}{Yao et~al\mbox{.}}{2022b}]%
        {ref:edge_cloud_survey}
\bibfield{author}{\bibinfo{person}{Jiangchao Yao}, \bibinfo{person}{Shengyu Zhang}, \bibinfo{person}{Yang Yao}, \bibinfo{person}{Feng Wang}, \bibinfo{person}{Jianxin Ma}, \bibinfo{person}{Jianwei Zhang}, \bibinfo{person}{Yunfei Chu}, \bibinfo{person}{Luo Ji}, \bibinfo{person}{Kunyang Jia}, \bibinfo{person}{Tao Shen}, {et~al\mbox{.}}} \bibinfo{year}{2022}\natexlab{b}.
\newblock \showarticletitle{Edge-Cloud Polarization and Collaboration: A Comprehensive Survey for AI}.
\newblock \bibinfo{journal}{\emph{IEEE Transactions on Knowledge and Data Engineering}} (\bibinfo{year}{2022}).
\newblock


\bibitem[\protect\citeauthoryear{Zhang, Kuang, Chen, Liu, Wu, and Xiao}{Zhang et~al\mbox{.}}{2022a}]%
        {zhang2022fairness}
\bibfield{author}{\bibinfo{person}{Fengda Zhang}, \bibinfo{person}{Kun Kuang}, \bibinfo{person}{Long Chen}, \bibinfo{person}{Yuxuan Liu}, \bibinfo{person}{Chao Wu}, {and} \bibinfo{person}{Jun Xiao}.} \bibinfo{year}{2022}\natexlab{a}.
\newblock \showarticletitle{Fairness-aware contrastive learning with partially annotated sensitive attributes}. In \bibinfo{booktitle}{\emph{The Eleventh International Conference on Learning Representations}}.
\newblock


\bibitem[\protect\citeauthoryear{Zhang, Kuang, Chen, You, Shen, Xiao, Zhang, Wu, Wu, Zhuang, et~al\mbox{.}}{Zhang et~al\mbox{.}}{2023b}]%
        {zhang2023federated}
\bibfield{author}{\bibinfo{person}{Fengda Zhang}, \bibinfo{person}{Kun Kuang}, \bibinfo{person}{Long Chen}, \bibinfo{person}{Zhaoyang You}, \bibinfo{person}{Tao Shen}, \bibinfo{person}{Jun Xiao}, \bibinfo{person}{Yin Zhang}, \bibinfo{person}{Chao Wu}, \bibinfo{person}{Fei Wu}, \bibinfo{person}{Yueting Zhuang}, {et~al\mbox{.}}} \bibinfo{year}{2023}\natexlab{b}.
\newblock \showarticletitle{Federated unsupervised representation learning}.
\newblock \bibinfo{journal}{\emph{Frontiers of Information Technology \& Electronic Engineering}} \bibinfo{volume}{24}, \bibinfo{number}{8} (\bibinfo{year}{2023}), \bibinfo{pages}{1181--1193}.
\newblock


\bibitem[\protect\citeauthoryear{Zhang, Feng, Kuang, Zhang, Zhao, Yang, Chua, and Wu}{Zhang et~al\mbox{.}}{2023a}]%
        {zhangsy2023personalized}
\bibfield{author}{\bibinfo{person}{Shengyu Zhang}, \bibinfo{person}{Fuli Feng}, \bibinfo{person}{Kun Kuang}, \bibinfo{person}{Wenqiao Zhang}, \bibinfo{person}{Zhou Zhao}, \bibinfo{person}{Hongxia Yang}, \bibinfo{person}{Tat-Seng Chua}, {and} \bibinfo{person}{Fei Wu}.} \bibinfo{year}{2023}\natexlab{a}.
\newblock \showarticletitle{Personalized Latent Structure Learning for Recommendation}.
\newblock \bibinfo{journal}{\emph{IEEE Transactions on Pattern Analysis and Machine Intelligence}} (\bibinfo{year}{2023}).
\newblock


\bibitem[\protect\citeauthoryear{Zhang, Jiang, Wang, Kuang, Zhao, Zhu, Yu, Yang, and Wu}{Zhang et~al\mbox{.}}{2020}]%
        {zhangsyDBLP:conf/mm/ZhangJWKZZYYW20}
\bibfield{author}{\bibinfo{person}{Shengyu Zhang}, \bibinfo{person}{Tan Jiang}, \bibinfo{person}{Tan Wang}, \bibinfo{person}{Kun Kuang}, \bibinfo{person}{Zhou Zhao}, \bibinfo{person}{Jianke Zhu}, \bibinfo{person}{Jin Yu}, \bibinfo{person}{Hongxia Yang}, {and} \bibinfo{person}{Fei Wu}.} \bibinfo{year}{2020}\natexlab{}.
\newblock \showarticletitle{DeVLBert: Learning Deconfounded Visio-Linguistic Representations}. In \bibinfo{booktitle}{\emph{{MM} '20: The 28th {ACM} International Conference on Multimedia}}. \bibinfo{publisher}{{ACM}}, \bibinfo{pages}{4373--4382}.
\newblock


\bibitem[\protect\citeauthoryear{Zhang, Liu, Zeng, Ooi, Tang, and Zhuang}{Zhang et~al\mbox{.}}{2023c}]%
        {zhang2023learning}
\bibfield{author}{\bibinfo{person}{Wenqiao Zhang}, \bibinfo{person}{Changshuo Liu}, \bibinfo{person}{Lingze Zeng}, \bibinfo{person}{Bengchin Ooi}, \bibinfo{person}{Siliang Tang}, {and} \bibinfo{person}{Yueting Zhuang}.} \bibinfo{year}{2023}\natexlab{c}.
\newblock \showarticletitle{Learning in Imperfect Environment: Multi-Label Classification with Long-Tailed Distribution and Partial Labels}. In \bibinfo{booktitle}{\emph{Proceedings of the IEEE/CVF International Conference on Computer Vision}}. \bibinfo{pages}{1423--1432}.
\newblock


\bibitem[\protect\citeauthoryear{Zhang and Lv}{Zhang and Lv}{2024}]%
        {zhang2024revisiting}
\bibfield{author}{\bibinfo{person}{Wenqiao Zhang} {and} \bibinfo{person}{Zheqi Lv}.} \bibinfo{year}{2024}\natexlab{}.
\newblock \showarticletitle{Revisiting the Domain Shift and Sample Uncertainty in Multi-source Active Domain Transfer}. In \bibinfo{booktitle}{\emph{Proceedings of the IEEE/CVF Conference on Computer Vision and Pattern Recognition}}.
\newblock


\bibitem[\protect\citeauthoryear{Zhang, Shi, Guo, Zhang, Cai, Li, Luo, and Zhuang}{Zhang et~al\mbox{.}}{2021}]%
        {zhang2021magic}
\bibfield{author}{\bibinfo{person}{Wenqiao Zhang}, \bibinfo{person}{Haochen Shi}, \bibinfo{person}{Jiannan Guo}, \bibinfo{person}{Shengyu Zhang}, \bibinfo{person}{Qingpeng Cai}, \bibinfo{person}{Juncheng Li}, \bibinfo{person}{Sihui Luo}, {and} \bibinfo{person}{Yueting Zhuang}.} \bibinfo{year}{2021}\natexlab{}.
\newblock \showarticletitle{MAGIC: Multimodal relAtional Graph adversarIal inferenCe for Diverse and Unpaired Text-based Image Captioning}.
\newblock \bibinfo{journal}{\emph{arXiv preprint arXiv:2112.06558}} (\bibinfo{year}{2021}).
\newblock


\bibitem[\protect\citeauthoryear{Zhang, Zhu, Hallinan, Zhang, Makmur, Cai, and Ooi}{Zhang et~al\mbox{.}}{2022b}]%
        {zhang2022boostmis}
\bibfield{author}{\bibinfo{person}{Wenqiao Zhang}, \bibinfo{person}{Lei Zhu}, \bibinfo{person}{James Hallinan}, \bibinfo{person}{Shengyu Zhang}, \bibinfo{person}{Andrew Makmur}, \bibinfo{person}{Qingpeng Cai}, {and} \bibinfo{person}{Beng~Chin Ooi}.} \bibinfo{year}{2022}\natexlab{b}.
\newblock \showarticletitle{Boostmis: Boosting medical image semi-supervised learning with adaptive pseudo labeling and informative active annotation}. In \bibinfo{booktitle}{\emph{Proceedings of the IEEE/CVF Conference on Computer Vision and Pattern Recognition}}. \bibinfo{pages}{20666--20676}.
\newblock


\bibitem[\protect\citeauthoryear{Zhang, Zhu, Song, Koniusz, King, et~al\mbox{.}}{Zhang et~al\mbox{.}}{2024}]%
        {zhang2024mitigating}
\bibfield{author}{\bibinfo{person}{Yifei Zhang}, \bibinfo{person}{Hao Zhu}, \bibinfo{person}{Zixing Song}, \bibinfo{person}{Piotr Koniusz}, \bibinfo{person}{Irwin King}, {et~al\mbox{.}}} \bibinfo{year}{2024}\natexlab{}.
\newblock \showarticletitle{Mitigating the Popularity Bias of Graph Collaborative Filtering: A Dimensional Collapse Perspective}.
\newblock \bibinfo{journal}{\emph{Advances in Neural Information Processing Systems}}  \bibinfo{volume}{36} (\bibinfo{year}{2024}).
\newblock


\bibitem[\protect\citeauthoryear{Zhou, Zhu, Song, Fan, Zhu, Ma, Yan, Jin, Li, and Gai}{Zhou et~al\mbox{.}}{2018}]%
        {ref:din}
\bibfield{author}{\bibinfo{person}{Guorui Zhou}, \bibinfo{person}{Xiaoqiang Zhu}, \bibinfo{person}{Chenru Song}, \bibinfo{person}{Ying Fan}, \bibinfo{person}{Han Zhu}, \bibinfo{person}{Xiao Ma}, \bibinfo{person}{Yanghui Yan}, \bibinfo{person}{Junqi Jin}, \bibinfo{person}{Han Li}, {and} \bibinfo{person}{Kun Gai}.} \bibinfo{year}{2018}\natexlab{}.
\newblock \showarticletitle{Deep interest network for click-through rate prediction}. In \bibinfo{booktitle}{\emph{Proceedings of the 24th ACM SIGKDD International Conference on Knowledge Discovery \& Data Mining}}. \bibinfo{pages}{1059--1068}.
\newblock


\bibitem[\protect\citeauthoryear{Zhu, Li, Shao, Hao, Wu, Kuang, Xiao, and Wu}{Zhu et~al\mbox{.}}{2023a}]%
        {DBLP:conf/mm/ZhuL0HWK0W23}
\bibfield{author}{\bibinfo{person}{Didi Zhu}, \bibinfo{person}{Yinchuan Li}, \bibinfo{person}{Yunfeng Shao}, \bibinfo{person}{Jianye Hao}, \bibinfo{person}{Fei Wu}, \bibinfo{person}{Kun Kuang}, \bibinfo{person}{Jun Xiao}, {and} \bibinfo{person}{Chao Wu}.} \bibinfo{year}{2023}\natexlab{a}.
\newblock \showarticletitle{Generalized Universal Domain Adaptation with Generative Flow Networks}. In \bibinfo{booktitle}{\emph{{ACM} Multimedia}}. \bibinfo{publisher}{{ACM}}, \bibinfo{pages}{8304--8315}.
\newblock


\bibitem[\protect\citeauthoryear{Zhu, Li, Yuan, Li, Kuang, and Wu}{Zhu et~al\mbox{.}}{2023b}]%
        {zhu2023universal}
\bibfield{author}{\bibinfo{person}{Didi Zhu}, \bibinfo{person}{Yinchuan Li}, \bibinfo{person}{Junkun Yuan}, \bibinfo{person}{Zexi Li}, \bibinfo{person}{Kun Kuang}, {and} \bibinfo{person}{Chao Wu}.} \bibinfo{year}{2023}\natexlab{b}.
\newblock \showarticletitle{Universal domain adaptation via compressive attention matching}. In \bibinfo{booktitle}{\emph{Proceedings of the IEEE/CVF International Conference on Computer Vision}}. \bibinfo{pages}{6974--6985}.
\newblock


\end{thebibliography}
